\documentclass[useAMS,usenatbib]{mn2e}
\usepackage{amssymb,amsmath,graphicx}
\usepackage{longtable}[1]
\usepackage{graphics}
\usepackage{lscape}
\usepackage{color}
\title[Star clusters in Mrk\,930]{Star cluster formation and evolution in Mrk\,930: properties of a metal-poor starburst\thanks{Based on observations made with the NASA/ESA Hubble Space Telescope, obtained at the Space Telescope Science Institute, which is operated by the Association of Universities for Research in Astronomy, Inc., under NASA contract NAS 5-26555. These observations are associated with program \# GO 10902.}}

\author[A. Adamo et al.]{A. Adamo$^{1}$\thanks{E-mail:
adamo@astro.su.se}, 
G. \"Ostlin$^{1}$,
 E. Zackrisson$^{1}$, 
 P. Papaderos$^{2}$, 
 N. Bergvall$^{3}$,  \newauthor
 R. M. Rich$^{4}$, 
 and G. Micheva$^{1}$
 \\
$^{1}$Department of Astronomy, Stockholm University, Oscar Klein Center, AlbaNova, Stockholm SE-106 91, Sweden\\
$^{2}$Centro de Astrofisica da Universidade do Porto, Rua das Estrelas 4150-762 Porto, Portugal\\
$^{3}$Department of Physics and Astronomy, Box 516, 751 20 Uppsala\\
$^{4}$Department of Physics and Astronomy, University of California at Los Angeles, \\
Physics and Astronomy Building 430 Portola Plaza, Box 951547 Los Angeles, CA 90095-1547}

\newcommand{\araa}{ARA\&A}
\newcommand{\apj}{ApJ}
\newcommand{\aj}{AJ}
\newcommand{\mnras}{MNRAS}
\newcommand{\aap}{A\&A}
\newcommand{\pasp}{PASP}

\newcommand{\apjl}{ApJ}
\newcommand{\msun}{M_{\odot}}
\newcommand{\ha}{H${\alpha}$}
\newcommand{\hb}{H${\beta}$}

\begin{document}

\date{Accepted xxxx yyyyy nn. Received xxxx yyyyy nn; in original form xxxx yyyyy nn}

\pagerange{\pageref{firstpage}--\pageref{lastpage}} \pubyear{2011}

\maketitle

\label{firstpage}

\begin{abstract}
 
We present the analysis of the large population of star clusters in the blue compact galaxy (BCG) Mrk\,930. The study has been conducted by means of a photometric analysis of multiband data obtained with the Hubble Space Telescope (HST). We have reconstructed the spectral energy distributions of the star clusters and estimated age, mass, and extinction for a representative sample. Similar to previous studies of star clusters in BCGs, we observe a very young cluster population with 70 \% of the systems formed less than 10 Myr ago. In Mrk\,930 the peak in the star cluster age distribution at 4 Myr is corroborated by the presence of Wolf-Rayet spectral features, and by the observed optical and IR lines ratios [O\,{\sc iii}]/\hb \,and [Ne\,{\sc iii}]/[Ne\,{\sc ii}].  The recovered extinction in these very young clusters shows large variations, with a decrease at older ages. 
It is likely that our analysis is limited to the optically brightest objects (i.e. systems only partially embedded in their natal cocoons; the deeply embedded clusters being undetected).
We map the extinction across the galaxy using low-resolution spectra and the  \ha/\hb \ ratio, as obtained from ground-based
narrow band imaging. These results are compared with the extinction distribution recovered from the clusters.  We find that the mean optical extinction derived in the starburst regions is close to the averaged value observed in the clusters (more than 80 \% of systems have $E(B-V) \leq 0.2$ mag), but locally, do not trace the more extinguished clusters. 
Previous HST studies of BCGs have revealed a population of young and extremely red super star clusters. We detect a considerable fraction of clusters affected by a red excess also in Mrk\,930. The nature of the red excess, which turns up at near-IR wavelengths ($I$ band and longward) remains unknown. 
We compare the cluster formation and the star formation history, the latter derived from the fit of spectral population synthesis models to  the spectra. We find a general agreement between the two independently estimated quantities. Using the cluster properties we perform a study of the host environmental properties. We find that the cluster formation efficiency (the fraction of star formation happening in clusters) is significantly higher, suggesting a key role of the environment for the formation of these massive objects.

\end{abstract}


\section{Introduction}

 Galaxy encounters give rise to vigorous bursts of star formation in which numerous of young star clusters are forming (e.g. \citealp{1995AJ....109..960W}, \citealp{1996AJ....112..416H}; \citealp{2000ApJ...532..845A}; \citealp{2008MNRAS.384..886V}; \citealp{2010AJ....139.1369P}). However, young star clusters are also observed in quiescent spiral galaxies, where the star formation rate (SFR) is lower and continuos (e.g. \citealp{1995AJ....110.1009B}; \citealp{1998AJ....115.1778C}; \citealp{2002AJ....124.1393L}; \citealp{2005A&A...431..905B}; \citealp{2009A&A...501..949M}). Massive star clusters form, during sporadic episodes, even in dwarf galaxies (NGC1569, \citealp{1985AJ.....90.1163A}; NGC1705, \citealp{1985A&A...149L..24M};  He2-10, \citealp{1994ApJ...423L..97C}; SBS 0335-052E, \citealp{1997ApJ...477..661T}; ESO338 IG-04, \citealp{1998A&A...335...85O};  \citealp{2002AJ....123.1454B}).

The luminous (M$_B< -18.0$), massive ($\sim 10^{9-10} \msun$) blue compact galaxies (BCGs) show clear signatures of interactions and/or mergers (e.g. \citealp{2001A&A...374..800O}) and the numerous observed massive clusters are likely formed in these encounters (\citealp{2003A&A...408..887O}; \citealp{A2010}; \citealp{A2010c}). The very bright ultraviolet and optical luminosities of these systems suggest rather low dust content and metallicity. Spectra dominated by emission lines clearly demonstrate that BCGs are undergoing a burst of star formation event. The youth of the burst episode is also observed in the recovered age distribution of the star clusters, which shows a peak of cluster formation younger than 5 Myr. BCGs are considered to be analogs of high redshift Lyman break galaxies and can probe galaxy formation and evolution at higher redshifts, when the Universe was younger and less enriched in metals. 

The analysis of the young cluster populations in BCGs is quite challenging due to the rapid evolution a cluster experiences during the first 10 Myr (still partly in an embedded phase). Moreover, this analysis is based on the integrated luminosities of the clusters, which are mostly unresolved at the distance of the targets.  Observations of resolved newly born star clusters in the Milky Way and in the Magellanic Clouds reveal that these are quite complex systems. A cluster  forms from the collapse and fragmentation of giant molecular clouds (GMCs; Lada \& Lada 2003). The compactness of the proto cluster determines whether the conglomerate of stars will form a cluster or a loose association (\citealp{2010ARA&A..48..431P}; \citealp{2010MNRAS.tmpL.168G}). The massive and short-lived stars rapidly reach the main sequence and produce strong winds and UV radiation, which ionize the intracluster gas and create bubbles and shells. These H{\sc ii} regions surround the optically bright core of stars and significantly contribute to the integrated fluxes. However, a large fraction of the stars is still accreting material from their dusty disks (young stellar objects, YSOs) or contracting (in the pre-main sequence; PMS phase). Simple stellar population models assume that PMS stars do not contribute to the integrated flux of the cluster. This assumption is valid at bluer spectral range but not in the IR. \citet{2005ApJ...630L.177M} estimated that PMS stars contribute between 5 and 17 \% to the total flux in the $H$ band during the first 3 Myr of cluster evolution. As follow up, \citet{2010ApJ...710.1746G}  found direct evidence of PMS stars in the spectrum of an unresolved young star cluster in the Antennae system. Moreover, the edges of the clusters are places for triggered \citep{1998ASPC..148..150E} and progressive star formation (e.g. \citealp{2007ApJ...665L.109C}; \citealp{2002AJ....124.1601W}). Delayed or triggered star formation processes in dense and dusty regions surrounding the cluster could explain a large fraction of massive YSOs contributing to the IR spectrum of a cluster a few Myr old. 

A significant fraction of young star clusters in Haro 11 \citep{A2010} and ESO 185-IG13 \citep{A2010c} shows a clear signature of a flux excess in the near-IR. The models we use \citep[][hereafter Z01]{Zackrisson et al.} include a self-consistent treatment of the photoionized gas, important during the first Myr of the cluster evolution \citep[e.g.][]{Krüger et al., Zackrisson et al., Anders & Fritze-Alvensleben, 2008ApJ...676L...9Z, R2009, A2010b}. \citet{A2010b} show that nebular emission non-negligibly affects the SEDs of the clusters during the first $10-15$ Myr of cluster evolution. In metal-poor environments, the contribution becomes smaller after 6 Myr on the blue side of the cluster spectrum, but lasts longer in the NIR wavebands, contributing between 10-40\% of the total NIR fluxes at $\sim$10 Myr. Since the models used in \citet{A2010, A2010c} already include a contribution from photoionized gas, the cause of the excess in BCGs star clusters resides in other mechanisms (e.g. YSOs, PMS stars, hot dust, etc.).

Mrk\,930 is a BCG located at roughly 72 Mpc\footnote{Value taken from NED, corresponding to a distance modulus of 34.27 mag and a recession velocity of 5485$\pm$10 km/s. These values are used in the present analysis.}. The galaxy was imaged and cataloged in the mid-60s and the 70s, during the first large-scale objective-prism survey for galaxies with blue and ultraviolet excess in their continuum radiation \citep{1986ApJS...62..751M}, i.e., the First Byurakan Spectral Sky Survey. Due to the low metallicity content ($12+\log($O$/$H$)=8.06$, \citealp{1998ApJ...500..188I}), Mrk\,930 was included in the late 90s among the galaxies studied for its similarity with high redshift Lyman break galaxies (LBGs). \citet{1998ApJ...500..188I} classified Mkr 930 as a Wolf-Rayet (WR) galaxy. Typically, WR signatures in the spectrum of a low metallicity galaxy is a rare phenomenon because the duration of the WR phase is shorter when the metallicity decreases (see \citealp{2000ApJ...531..776G} and references therein). The presence of WR features indicates that  the galaxy is undergoing a young burst episode. 

Mrk\,930 was included in Malkan's {\it Hubble Space Telescope} survey of local active galaxies \citep{1998ApJS..117...25M}, as a H{\sc ii} galaxy. The short F606W exposure of the galaxy revealed that the two starburst knots observed in ground-based images were in reality formed by numerous bright star clusters \citep{2000ASPC..211...63O}. The presence of a very active starburst episode, the low metallicity environment, and the numerous clusters, make Mrk\,930 galaxy of great interest  to include in our statistical analysis of star cluster populations in extreme environments (\citealp{2003A&A...408..887O}; \citealp{A2010}, \citealp{A2010c}).

\begin{figure*}
\resizebox{0.7\hsize}{!}{\rotatebox{0}{\includegraphics{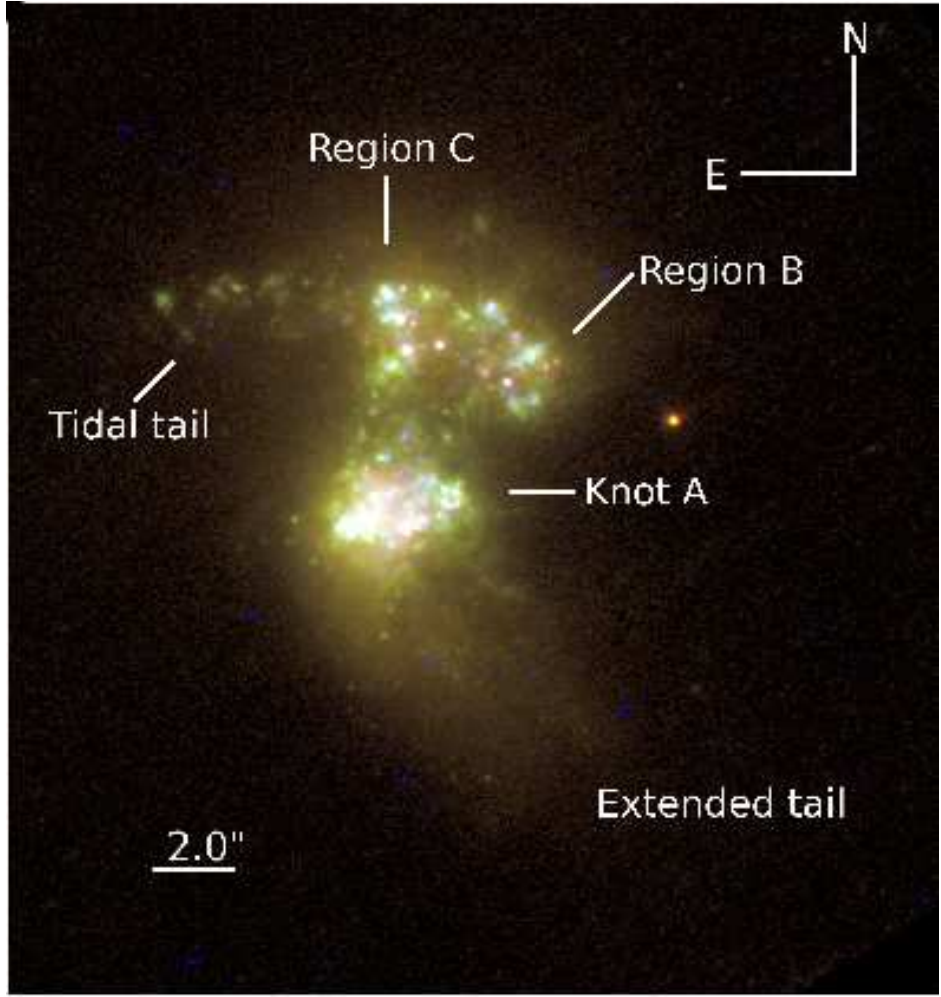}}}
\caption{Three-color image of the starburst galaxy Mrk\,930: {\it WFPC2} F814W filter in red, F606W in green and F336W in blue.  Some intense blue spots are caused by an imperfect removal  of cosmic rays from the WFPC2/F336W frame. Different star forming regions can be observed in the galaxy. The youngest are surrounded by intense nebular emission (green diffuse light, due to the wide V band filter F606W which transmits [O\,{\sc iii}] and H$\alpha$). Knot A is the brightest optical and UV region in the galaxy and is undergoing the most active starburst event. Region B and C are less clustered and coeval to knot A. The color of the extended tail in the South of the galaxy is redder than the starburst regions, probably formed by a more evolved stellar population. See the main text for more details.}
\label{h11b}
\end{figure*}

A 3-color composition of Mrk\,930, obtained with the high-resolution {\it HST} Planetary Camera (Figure~\ref{h11b}), shows a morphologically perturbed galaxy. Nebular emission surrounds the burst regions (the green regions are caused by strong nebular emission lines falling in the band pass of the wide V band filter F606W), confirming the youth of the stellar population. Knot A is the brightest region in the galaxy. It is also the most rich in star clusters, resembling the cluster complexes observed in M82 \citep{2006MNRAS.370..513S}  and Haro 11 \citep{A2010}. In the Southern part, an extended tail is observed.  The color of the tail is redder than the main galactic body, suggesting an older stellar population. Such an evolved population has also been detected in the outskirts of the galaxy (Micheva et al. in prep.). The northern part of the galaxy is formed by two close cluster-rich regions B and C and a tidal tail. Optical Fabry-Perot interferometric imaging shows signatures of a recent merger event in the velocity field of the galaxy (Marquart et al. in prep. 2011). 

Here, we present a complete multiwavelength analysis of the cluster population in Mrk\,930. The data are described in Section \ref{data-sample}. The cluster analysis and derived properties are discussed in Section \ref{prop_sc}. In this section we propose possible scenarios to explain the origin of the red excess. In Section \ref{ciao} we map the mean extinction distribution in the galaxy, using ground-based spectroscopy and imaging. We compare this mapping with the extinction derived locally from the clusters. In the following Section \ref{ciao2}, we discuss the properties of the galaxy as derived from the cluster formation and star formation history (CFH and SFH, respectively). Conclusions are summarized in the final section.

Throughout the paper we will use the Vega magnitude system.

\section[]{The data}
\label{data-sample}
\subsection{HST data: Observations \& Photometric analysis}
\label{ciao3}

{\it HST} multiband high-resolution imaging of MRK\ 930 were carried out in 2007 (associated with program \# GO 10902, PI:
G. \"Ostlin.). The galaxy was sampled with the same data set as for the ESO 185-IG13 target \citep{A2010c}. The ACS solar blind channel (SBC) camera was used to image the galaxy in the $FUV$ filter, F140LP. The optical sampling of the galaxy was originally scheduled for ACS/HRC and WFC but due to the failure of the instrument in 2007 we switched to WFPC2 and in particular  the Planetary Camera (PC) aperture and obtained mages in the $U$ (F336W), $B$ (F439W), $R$ (F606W), and $I $  (F814W) bands . Finally, the IR ($H$ band, F160W) imaging was performed by the NICMOS3 camera, which offers lower spatial resolution but wider field of view. 
The frames were reduced, drizzled and aligned using the {\tt MULTIDRIZZLE} task
(\citealp{2002hstc.conf..337K}; \citealp{2002PASP..114..144F}) in {\tt
PyRAF/STSDAS}\footnote{STSDAS and PyRAF are products of the Space
Telescope Science Institute, which is operated by AURA for NASA}. The {\it WFPC2}  F606W and F814W, the SBC/F140LP and the NIC 3  imaging were achieved using a dither pattern to improve the final resolution of the science frames. In the case of the F336W and F439W imaging, two exposures each were taken to perform cosmic ray correction, but not dithering was applied due to the shorter available exposure time. The final UV and optical frames were rescaled to $0.025 "/pixel$. The best resolution we could achieve for the NIC3 final frame was of  $0.067"/pixel$. Aperture photometry on the sources was done using a radius of $0.1"$. The sky annulus around the source was placed at 0.125" and had a width of 0.05". Due to the crowding of the regions, we preferred to use the same aperture radius and sky annulus in all the frames and correct for the missing flux (caused by a fixed aperture), using estimated aperture correction values for each frame (see Table \ref{table-obs}). A detailed description of the data reduction, cluster catalogue extraction, point source photometry and charge transfer efficiency correction is given in \citet{A2010, A2010c}. A list of the filters, total exposure time, zero points (ZPs), and other observational properties related to the reduction of the data are summarized in the Table \ref{table-obs}.

\begin{table*}
  \caption{{\it HST} observations carried out within the framework of program \# GO 10902 (PI G. \"Ostlin). For each filter we list in the table: the total exposure time; the corresponding ZPs  (Vega magnitude system); the aperture corrections, a$_c$; the number of objects detected with a $S/N\geq5$; and the corresponding magnitude limit. $a$) Both the {\it HST} filter names and the abbreviated  nomenclature indicated in bracket are used thereafter in the text.}
\centering
  \begin{tabular}{|c|c|c|c|c|c|c|c|}
  \hline
  Instrument &Filter$^{a}$ & Camera &  Exposure time & ZP (mag)&a$_c (mag)$ &N($\sigma\leq 0.2$)&  mag limit\\
   \hline
 ACS&F140LP ($FUV$)&SBC&2532 s&20.92&-0.54$\pm$0.05& 117 &24.0\\
  \hline
WFPC2& F336W ($U$)&PC&1200 s&19.43&-0.42$\pm$0.05&81&23.0\\
WFPC2& F439W ($B$)&PC&800 s&20.88&-0.36$\pm$0.05&79&24.1\\ 
WFPC2& F606W ($R$)&PC&4000 s&22.89&-0.61$\pm$0.09&207&27.1\\
WFPC2& F814W ($I$)&PC&4500 s&21.64&-0.73$\pm$0.04&207&26.0\\ 
\hline
 NICMOS&F160W ($H$)& NIC3 &4992s&21.88&-2.45$\pm$0.37&78&25.9\\
  \hline
\end{tabular}
\label{table-obs}
\end{table*}

\begin{figure}
\resizebox{\hsize}{!}{\rotatebox{0}{\includegraphics{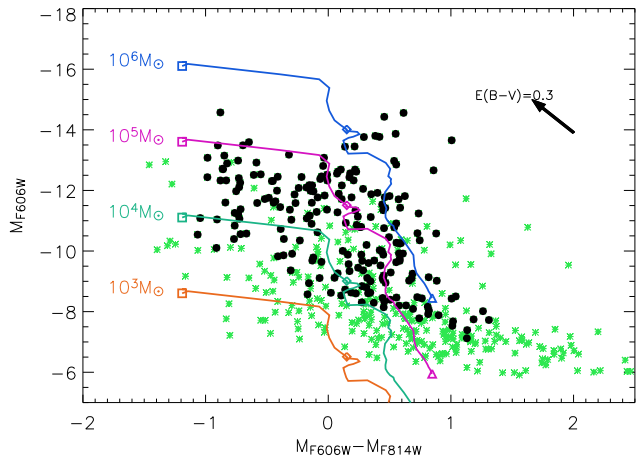}}}
\caption{Color magnitude diagram of the cluster population. Top panel: the small green (grey) asterisks show the 477 cluster candidates detected in the galaxy. The black filled dots are the 207 clusters with photometric errors in $I$ and $R$, $\sigma_m \leq 0.2$. The extinction vector indicates the shift of a cluster in this plot, if corrected for E(B-V)=0.3 mag. The evolutionary track from Z01 are for Z=0.004 and several values of the mass.  In each model track, the squares, diamonds, and triangles indicate 1 Myr, 10 Myr, and 14 Gyr, respectively. See main text for details.}
\label{CMD}
\end{figure}

As already found in \citet{A2010, A2010c} the number of clusters detected in the two deepest exposures, $R$ and $I$ bands, is greater than in the other available filters (see Table \ref{table-obs}). In total we recovered 477 cluster candidates (with $\sigma \leq 1.0$ mag). Viewing the color-magnitude diagram (CMD; Figure~\ref{CMD}) and evolutionary tracks we expect the cluster population to show a wide range of masses and ages. 
 After a photometric error cut at $\sigma \leq 0.2$ mag, 207 objects remain. Of these 157 sources are detected in at least a third filter, so this is the number of objects we consider in the analysis of Mrk\,930\footnote{The complete photometric catalogue is available on request from the
authors.}. The final photometric catalogue contains cluster candidates that have been detected in at least three filters with a signal-to-noise ratio of 5 or larger (i.e., photometric error of $\sigma_m \leq 0.2$). In Table \ref{table-obs} we show the corresponding magnitude limits and the number of objects retained in each frame after this selection. The final error associated with the integrated flux in each filter is the combined quadrature of the photometric error and the uncertainty on the aperture correction determination.  The foreground galactic extinction \citep{1998ApJ...500..525S} correction is also included.

\begin{figure}
\resizebox{\hsize}{!}{\rotatebox{0}{\includegraphics{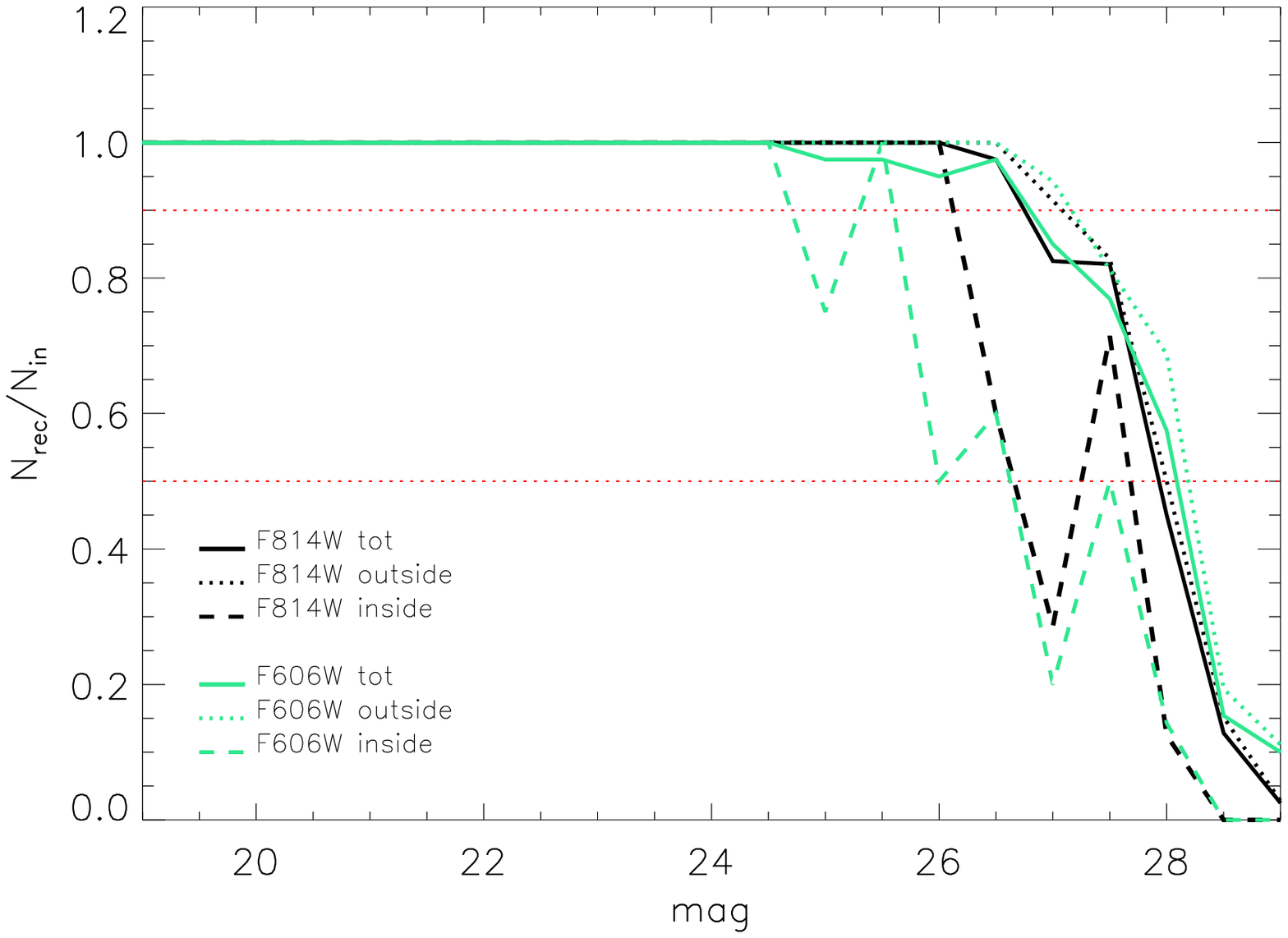}}}
\caption{Completeness fraction as function of the object magnitudes. The outputs of the F606W frame are plotted in green(grey), while in black we show the fraction for the F814W frame. See the main text for the definition of "inside" and "outside" regions. The red dotted straight lines show the 90 \% and 50 \% completeness limits.}
\label{compl_test}
\end{figure}

The completeness test was performed on the two $R$ and $I$ {\it WFPC2} frames, the deepest filters used to extract the first cross-correlated source catalogue \citep{A2010, A2010c}. This type of analysis is useful for constraining the sensitivity of our method to recover objects. Similar to the Haro 11 and ESO 185-IG13 analysis, we defined a center of the galaxy ($\alpha = 23:31:58.5$; $\delta = 28:56:52.0$) and a crowded region of 4.5" radius, where most of the starburst knots were contained. The surrounding region (an annulus of inner radius of 4.5" and outer radius of 12.25") is characterized by a low sky background, ideal for detecting faint objects. The number of recovered objects was compared to the crowding of the considered region. In Figure~\ref{compl_test}, we show the total fraction of recovered detections. The two solid lines (green and black) show that in general we are able to detect most (90 \%) of the objects with luminosity brighter of $\sim 26.8$ mag.  At roughly 28 mag we are still able to detect 50 \% of the objects in both filters. However this is an averaged result. Looking at the recovered fraction in the "inner" (crowded) and "outer" (starburst outskirts) areas, one can see that the detection of objects drops more rapidly in the crowded regions and changes quite abruptly. This behaviour is caused by the small area of the starburst region which does not allowed to build a statistically populated sample of objects at different luminosities bins. Since the quality and sensitivity of the used data for the three BCGs are similar (same exposure time and observational strategy), we can use the results of the completeness tests of Haro 11 and ESO 185-IG13 to consider approximately 26 mag as a threshold for 90\% completeness detection in the inner region for both filters.

\subsection{Ground based imaging and spectroscopy}

\subsubsection{NOT Narrow-band \ imaging}

Ground based narrow band imaging was carried out with the 2.56m Nordic Optical Telescope (NOT) on La Palma with the ALFOSC instrument. The H$\alpha$
observations were carried out in Sept 2002.
We obtained 1800s of integration through NOT filter \#70 approximately
centered on H$\alpha$ at the redshift of Mrk \,930. For continuum subtraction
we obtained 2400s of integration through filter \#78. We also obtained exposures
in both filters of  the spectrometric standard stars BD+28d4211 and GD248.
The data were reduced in a standard manner (bias, flatfielding, zeropoint calibration).
To obtain the H$\alpha$ flux in the continuum subtracted image we applied
the following additional steps: i) We corrected for the relative transmission of
the filter at the redshifted wavelength of H$\alpha$. ii) We corrected for the contribution of [N\,{\sc ii}] by taking the relative filter transmission at the redshifted wavelengths into account and by using spectroscopically determined line
ratios which we assumed to have a spatially constant value over the face of
the galaxy  (this is reasonable due to the similar ionization potentials). iii)
We multiplied with the equivalent rectangular width (the integral of the filter throughput curve divided by the filter peak transmission) of the filter to obtain
an image calibrated in units of erg/s/cm$^2$ .

The H$\beta$ observations were carried out in August 2007, using filters
\#113 ($t_{\rm exp}=4900$s) to capture H$\beta$ and \#17 (Str\"omgren-$b$,
$t_{\rm exp}=2100$s) for the continuum, and the spectrophotometric standard Feige\,110 for calibrations. The same reduction and calibration steps as for
H$\alpha$ were followed except that we did not need to correct for [N\,{\sc ii}]. 

In addition we obtained exposures in filters \#66 (600s) and \#18 (Str\"omgren-$y$,
600s) to capture the redshifted [O\,{\sc iii}]$_{\lambda 5007}$ line and its continuum.
Seeing was close to one arcsecond for all observations and conditions were photometric.

\subsubsection{Spectroscopy}

Low-resolution long slit spectra of Mrk\ 930 along two position angles (PAs) were taken on 
November 14, 2009 with the 3.6m Telescopio Nazionale Galileo (TNG) using the  
DOLORES focal reducer. We used the LR--B grism which yields a wavelength
coverage between 3000 and 8430 \AA and a resolution R$\sim$600 for our slit
width of 1\arcsec. The data were acquired in 1$\times$2 pixel binning,
resulting in a spatial scale of 0\farcs504 per pixel. 
Several spectroscopic standard stars were observed at various airmasses (AM) 
for the sake of flux calibration. 
The observations consisted of three consecutive 12 min exposures along each
position angle and were carried out in either case at a low airmass ($\leq$1.03). 
No corrections for differential atmospheric refraction were therefore applied.

Conditions were relatively good during the first pointing with the slit rotated to 
PA=74.1 deg and encompassing regions B and C; comparison of multiple science
and calibration exposures suggest transparency varied by $<$5\%. 
Despite the mediocre seeing (1\farcs5 to 1\farcs9 FWHM), we could
clearly separate  regions B and C.   
For the other slit position the seeing was worse,  $\approx$2\farcs5, 
rendering a spatially resolved study of component A (PA=108 deg) impossible. 
An intercomparison of individual galaxy and standard star
spectra taken during that phase of the observing run has revealed transparency 
variations by a factor of two, resulting in a significantly lower signal-to-noise (S/N)
ratio for at least one of the three spectra taken. 
The data were reduced with MIDAS using a standard procedure.

\section{Extracting the cluster properties from their luminosity}
\label{prop_sc}

\subsection{The Cluster Luminosity Functions}

It is known that young star clusters have power law luminosity (CLF, $dN(L) \propto
L^{-\alpha} dL$) and mass functions (CMF, $dN(M)=CM^{-\gamma}$ dM), with slopes of $\sim 2.0$. However, the CLF varies from galaxy to galaxy, showing a wide range of slopes, $1.8 \leq \alpha \leq 2.4$ (\citealp{2002AJ....124.1393L}; \citealp{2003AJ....126.1836H}; \citealp{2003MNRAS.343.1285D}; \citealp{2009A&A...501..949M}; \citealp{2010AJ....139.1369P}; \citealp{2010AJ....140...75W}; among many others), or even a bend occurring roughly between $-8.0$ and $-10.0$ mag (Gieles et al. 2006a,b).

 The CLF is composed by many cluster populations, formed in different periods of the star formation history of the host. Absence or presence of variations in the slope of the CLF could probe: i) whether the CMF is a single power law with index $-2.0$ (see \citealp{2010ApJ...719..966C}, \citealp{2010AJ....140...75W}) or has a Schechter function with a "soft" truncation at a characteristic mass, related to the environmental properties of the host galaxy (see \citealp{2006A&A...450..129G}, \citealp{2008MNRAS.390..759B}, \citealp{2009A&A...494..539L}), ii) whether cluster disruption is mass dependent or independent (see \citealp{2009Ap&SS.324..183L} for a short summary).

\begin{figure*}
\resizebox{0.33\hsize}{!}{\rotatebox{0}{\includegraphics{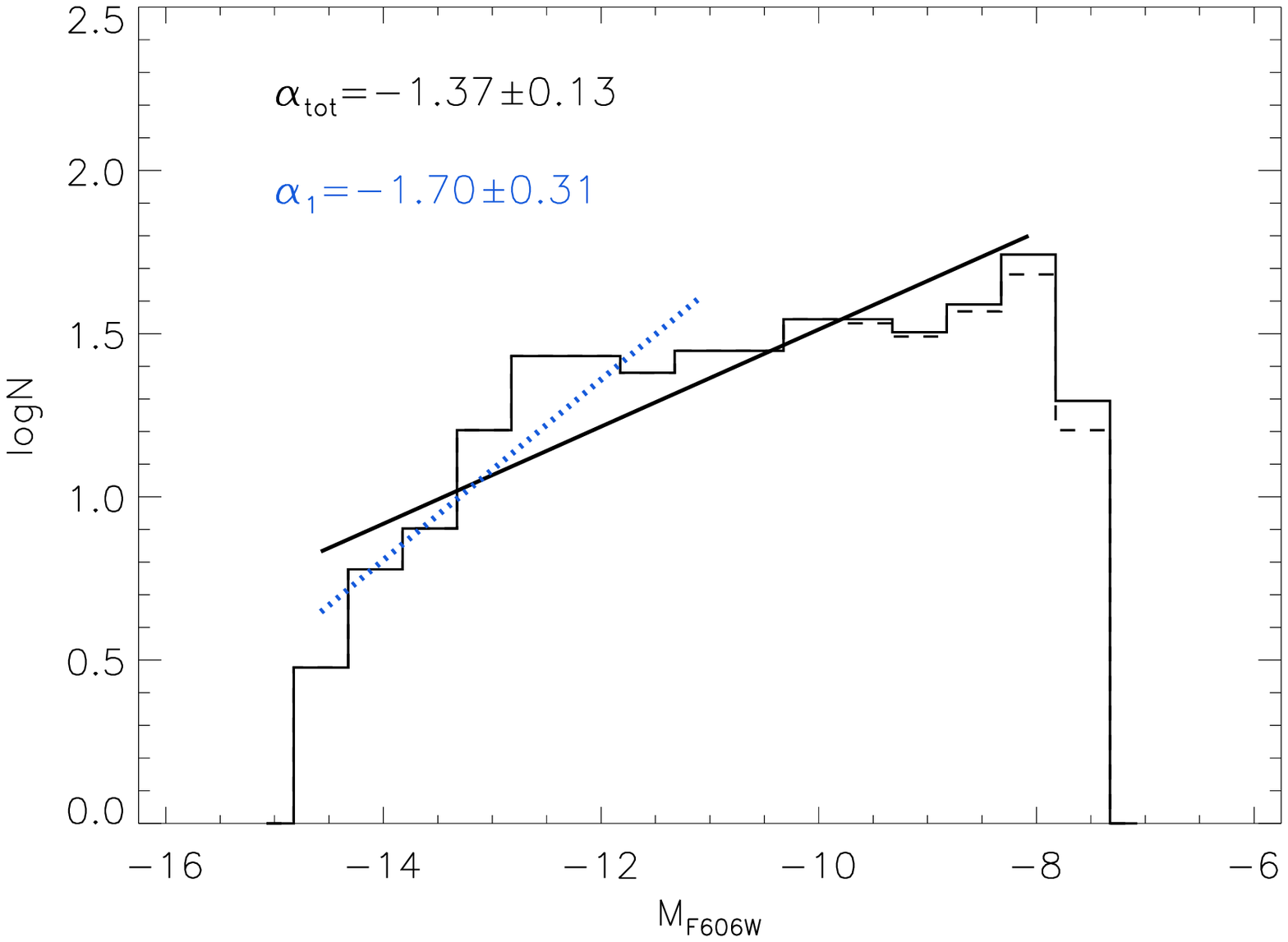}}}
\resizebox{0.33\hsize}{!}{\rotatebox{0}{\includegraphics{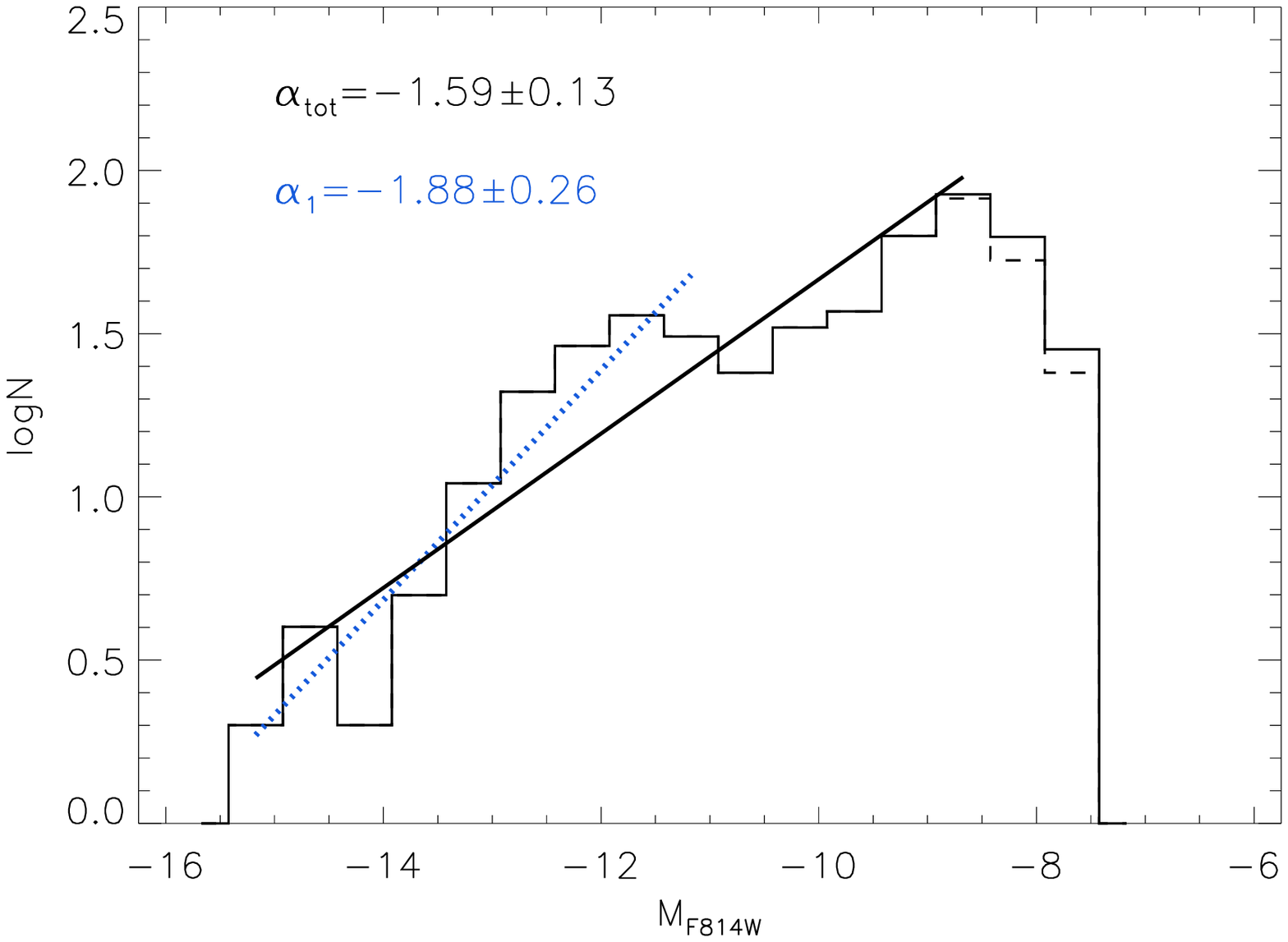}}}
\resizebox{0.33\hsize}{!}{\rotatebox{0}{\includegraphics{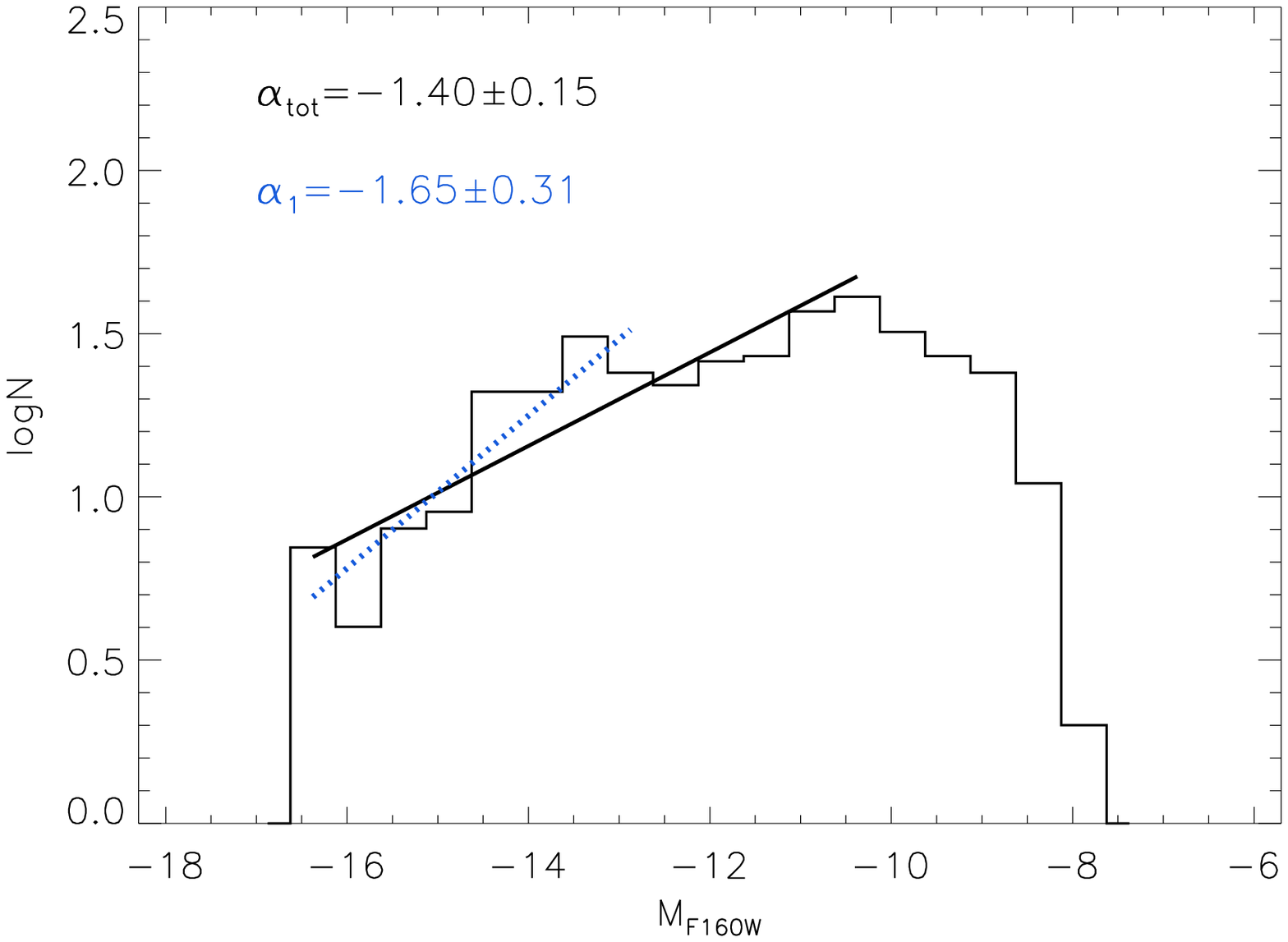}}}
\caption{Recovered cluster luminosity functions in three of the six filters available in this analysis. The histograms in solid black line are the distributions corrected for completeness. The underlying dashed histograms show the initial distributions. In the right panel, the histogram represents  the recovered distribution. No completeness correction is available for this filter. The fit to different luminosity bins are showed in different colors (line styles). The resulting slopes are included in the insets.}
\label{clf}
\end{figure*}

In studies of star cluster properties in BCGs it has been argued that the CLF in these systems is shallower than expected: $\alpha \sim 1.8$ in ESO 338-IG04 \citep{2003A&A...408..887O};  $\alpha \sim 1.5$ (but $\alpha \sim 1.6$ if only the brightest bins are included in the fit) in Haro 11 \citep{A2010}; $\alpha \sim 1.8$ (but $\alpha \sim 2.2$ including only the brightest bins) in ESO 185-IG13 \citep{A2010c}. 

We present here the analysis of the CLF in Mrk\,930. We use the numerous objects detected in the two deepest exposures, F606W and F814W, with photometric errors below 1.0 mag, to determine the slope of the CLF (similar results have been recovered if only the $\sigma \leq 0.2$ sample is used). In the left and central panels of Figure~\ref{clf}, we show the two recovered CLFs corrected for completeness. In the right panel we include the CLF for the $H$ band. In this last case, it is important to mention that we included all the positive detections obtained, making the derived distribution rather uncertain.

The fit to the $R$ and $I$ CLFs, showed as a black solid lines, includes all the luminosity bins up to the value where we recovered 90 \% of objects in the completeness tests. The fit to the $H$ band CLF excludes the fainter bins where we notice a decrement in the number of objects. The slopes in the three filters are, inside the uncertainties, similar and much flatter than expected. A second fit, confined to the brightest luminosity bins, produces higher values, slightly more consistent with the expected 2.0 value.

We have already discussed in \citet{A2010, A2010c} that blending can flatten the observed CLF. Moreover, the incompleteness could be much more severe than the values obtained in this analysis. Therefore, it is not possible to derive any conclusion with the current data. Simulations, which reproduce blending effect and different CMF shapes, need to be used to explore the physical properties of the host environments by means of the CLF.

\subsection{Analysis method: $\chi^2$ fitting}
\label{chi2}

To extract the physical properties of the clusters, i.e., age, mass, and extinction, we compared spectral synthesis models to their reconstructed spectral energy distributions (SEDs) from multiband photometry. 
The models we used contain several assumptions and parameters:
\begin{description}
\item i) clusters form in an instantaneous burst, i.e., the stellar population is coeval;
\item ii) the stars in the cluster have masses sampled by a Kroupa  universal initial mass function \citep{2001MNRAS.322..231K};
\item iii) both photoionized gas (important during the first 10 Myrs of the cluster evolution) and stars contribute to the total integrated fluxes;
\item iv) standard H\,{\sc ii} region values for the gas as input parameters to Cloudy (version 90.05, \citealp{Ferland et al.}): hydrogen gas density of $10^{2}$ cm$^{-3}$, filling factor of 0.01, covering factor of 1 (there is no-leakage of ionizing photons, e.g. all the Lyman continuum photons ionize the gas);
\item v) the metallicity of the stars and the gas in the cluster are the same, i.e., $Z=0.004$ (corresponding to the values determined by \citealp{1998ApJ...500..188I}).
\end{description}

A detailed description of the models can be found in Z01  and in \citet{A2010, A2010c}. The spectral
synthesis models of \citet[][hereafter M08]{Marigo et al.} have been used to test the robustness of
some of our results. 

To constrain the model which most closely reproduced the observed cluster SED we performed a $\chi^2$-fit. Applying the Calzetti attenuation law \citep{2000ApJ...533..682C}, we created  a grid of models with values of extinction, $E(B-V)$, ranging from 0.0 to 3.0 for each age step. The reddened model with the smallest weighted residuals was considered the best one and  produced age and extinction for the cluster. The mass was proportional to the normalization factor between the best model and the observation. A detailed description of the $\chi^2$-algorithm is given in \citet{A2010}.

\subsection{Reconstructing the SEDs of the clusters}
\label{SED_par}
A close look at the $\chi^2$-fit outputs reveals that also in Mrk\,930 there is a fraction of clusters with a NIR excess \citep{A2010, A2010c}, i.e., an observed luminosity at wavelengths larger than 0.8 $\mu$m which cannot be reproduced by our models. The fit procedure for clusters affected by a red excess and their analysis has widely been discussed in \citet{A2010, A2010c}. Here, we summarize the main characteristics. 

In sources affected by a red excess the fit to the whole SED (from UV to IR) fails, e.g. producing high residuals and wrong estimates of the cluster age (mass). To trace which of the clusters are affected by a red excess, two new fits are performed. In these two runs we exclude once the IR band ($H$ band) and in the other the $I$ and $H$ bands (the latter one if available). The requirement of at least 3 data points available, in order to perform the fit, is always retained. The outputs of the three fits are then compared and, for each cluster, we consider as best fit the one which better reproduces the UV and optical shape of the observed cluster SED. In general, we notice that even if the I band excess is less evident than the IR one and could be easily overlooked, it has an important impact on the estimated ages and masses \citep[see][]{A2010, A2010c}. 

In Mrk\,930 we performed a $\chi^2$ analysis on a total of 157 objects. Of these objects,  61 \% have normal SEDs, and, inside the photometric uncertainties, the performed analysis produced fairly good fits. 12 \% of clusters display an $H$ band flux excess (IR excess) and the remaining 22 \% have a flux excess at $\lambda \geq $0.8  $\mu$m (NIR excess). For these clusters, the age, mass, and extinction have been estimated from the UV and optical SEDs. In a small number of cluster candidates (7 in total, $\sim$5 \%), the fit procedure failed therefore they are excluded from further analyses.

Finally, we carried out an individual analysis for a rather bright and isolated point-like system, whose nature remains unclear. This object is easily noted in Figure~\ref{h11b} on the right (west direction) side of Region B, located at a projected distance of roughly 1 kpc. It looks reddish, and different from the color of the clusters in the starburst regions. Interestingly, it is not detected in the $FUV$ and $U$ bands and is only marginally detected in the $B$ band. Different scenarios can be advocated to explain the nature of this object: either a deeply embedded and very young cluster (similar cases have been found in ESO 185-IG13, \citealp{A2010c}), or a background object (early type galaxy at redshift $\sim 0.1$ or Lyman break system at redshift $>$ 3.0). 

Since it is isolated we performed  a new photometric analysis, using an aperture radius of 1.0". The model fit to the SED of this cluster candidate produced an age of 35.0 Myr, a mass of $1.2\times10^7 \msun$, and $E(B-V)=0.68$ ($A_V \sim 2.5$ and $A_B \sim 3.45$ mag). The properties constrained for this object are quite extreme. The mass is similar to the values found for W3 and W30 in the merger remnant NGC 7252, and the globular cluster G114 in NGC 1316 (\citealp{2004A&A...416..467M}; \citealp{2006A&A...448..881B}). These evolved clusters are located a few kiloparsecs away from their hosts, have ages between 400 Myr and 3 Gyr, and dynamical masses between $1-8 \times 10^7 \msun$. The formation mechanism for such massive systems is unclear. Numerical simulations \citep{2005MNRAS.359..223F} suggest the possibility for these systems to form from merger of super star clusters born inside the same cluster complex. 

The cluster-like object in Mrk\,930 is much younger  than these super massive evolved clusters, and even if it potentially could be a predecessor of such rare objects, we cannot exclude any scenario with the current data (i.e., background object). The extracted spectrum at the position of this cluster shows a few faint emission lines (e.g., H$\alpha$ and H$\beta$) at the same rest frame position of the galaxy. The continuum has a low S/N and disappeared around 4200 \AA, in agreement with the reconstructed SED from the photometric data. However, we note that locally the region surrounding the cluster has similar nebular emission lines suggesting contamination from the galaxy.  Due to the uncertain nature of this cluster like object, we exclude it from the cluster analysis. 

\subsection{Cluster photometric properties}

\begin{figure*}
\resizebox{0.48\hsize}{!}{\rotatebox{0}{\includegraphics{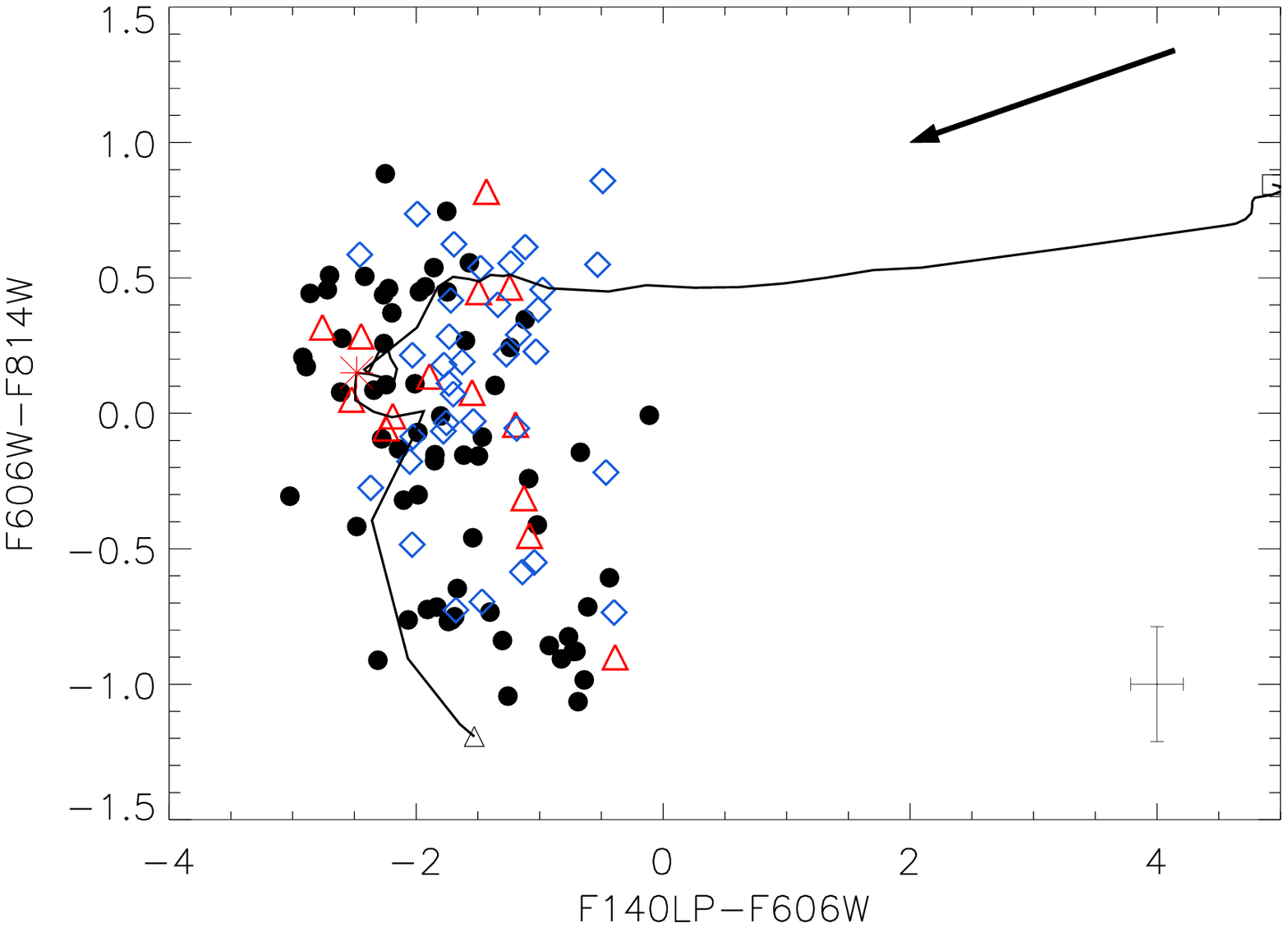}}}
\resizebox{0.48\hsize}{!}{\rotatebox{0}{\includegraphics{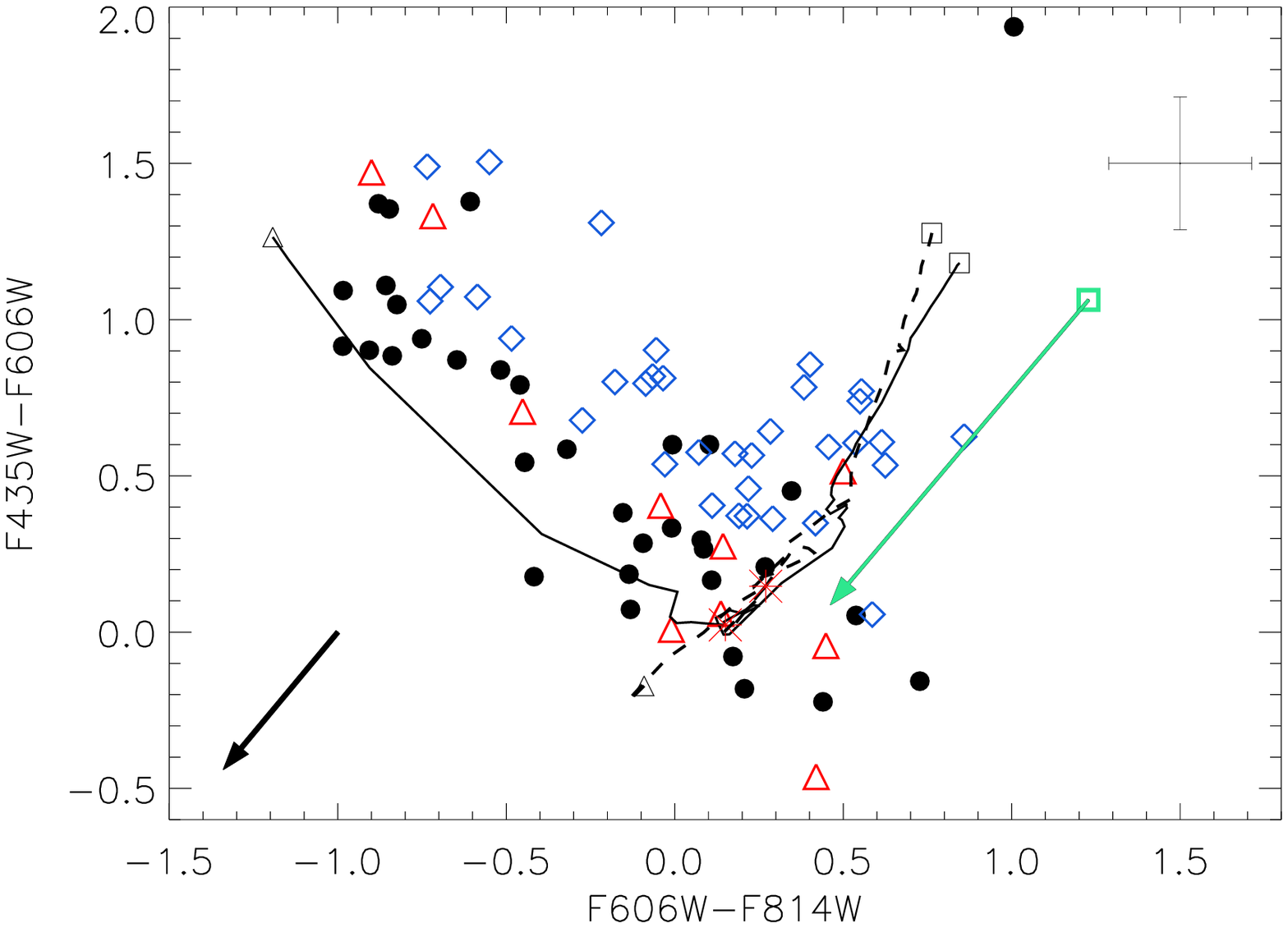}}}\\
\resizebox{0.48\hsize}{!}{\rotatebox{0}{\includegraphics{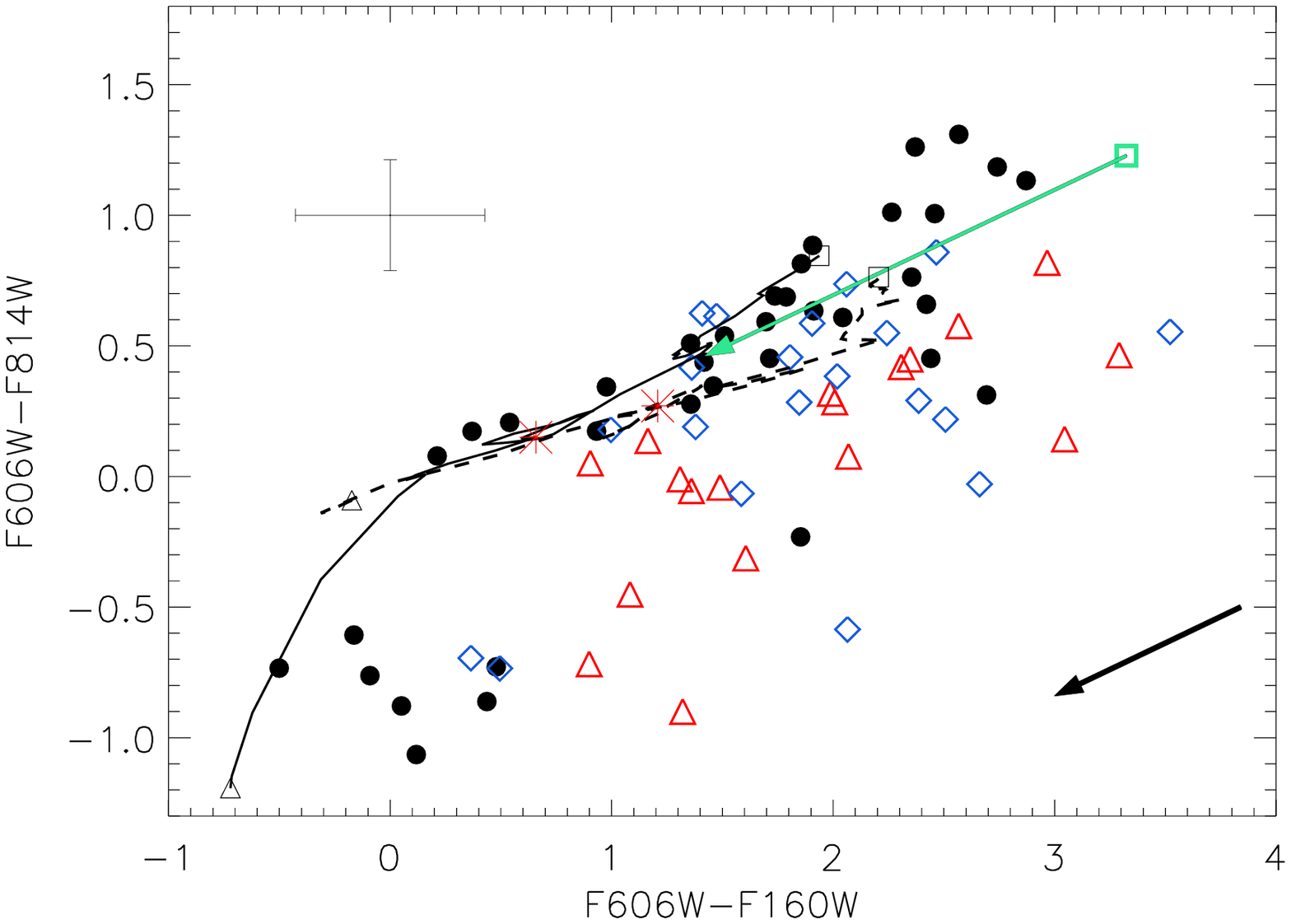}}}
\caption{Color-color diagrams of the cluster population in Mrk\,930. Different filter combinations are compared to the $R-I$ color (F606W-F814W). Upper left: $UV-R$ (F140LP-F606W); Upper right: $B-R$ (F435W-F606W); Center: $R-H$ (F606W-F160W). In each panel, the Z01 evolutionary tracks are plotted as a solid black line. Where  predictions for the used filters were available we included the M08 tracks (dashed lines) as well. The black triangles show the starting point of the tracks (1 Myr for Z01 and 4 Myr for M08), the red asterisks the position in the tracks at 10 Myr, and the black squares at 14 Gyr. The black dots are clusters with fully fitted SEDs (from UV to IR). The red triangles are cluster with an excess at $\lambda >1.0$ $\mu$m. The blue diamonds shows objects with an excess starting longword $\lambda \geq 0.8$ $\mu$m. The black arrows indicate in which direction and quantity the colors of the clusters change if a $E(B-V)=0.3$ is applied. Mean errors are included. The green square shows the position of the cluster-like system in the color-color diagram, and the green arrow the position after an extinction correction of $E(B-V)=0.68$ is applied (see Section \ref{SED_par}).}
\label{ccd}
\end{figure*}

Color-color diagrams  represent useful tools for analysing the numerous Mrk\,930 cluster population.  In Figure~\ref{ccd} we plot 3 different colors against $R-I$, which is used as reference. Generally, a negative $R-I < 0$ color indicates ages younger than 10 Myr and is produced by a strong nebular contribution to the $R$ filter, which includes the bright \ha \, line. However, this assumption is not always valid due to the flux excess in the $I$ band. The clusters, in the diagrams, have different symbols depending of their observed SEDs: black dots indicate normal SEDs, red triangles clusters with an excess at $\lambda \geq 1.0$ $\mu$m (hereafter, IR excess), and blue diamonds SEDs which deviate at $\lambda \geq 0.8$ $\mu$m (hereafter, NIR excess). We notice that clusters with a NIR excess have a $R-I$ color between 0.2 and 1.2 mag redder than the prediction made by the best SED fitting model. These residuals can be observed in Figure~\ref{delta-exc} and will be discussed in Section \ref{rex-cause}.  The $R-I$ color of the "blue" cluster-diamonds is such that, if overlooked, causes age (and mass) overestimates, affecting the results of the optical based cluster analysis.   

The $UV-R$ (upper left panel) color shows that the clusters detected in the $FUV$ are young, at least younger than 30-40 Myr. Many of the clusters with an NIR excess are located in an area where $R-I > 0$, e.g., with ages older than 10 Myr. The $FUV$ band is sensitive to the reddening. Looking at the color-color diagram one can see that clusters detected in the $FUV$ have extinctions, $E(B-V) \leq 0.3$ ( e.g., the arrow in the plot). Therefore, we consider these very young $FUV$ bright clusters as systems which have already gone through the deeply embedded phase. 

A combination of optical colors (upper right panel) probes the impact of the contribution from photoionized gas. We compare our models (solid black line) to the Padova tracks (dashed black line; \citealp{Marigo et al.}). These last models offer a better modelling of the asymptotic giant brach (the difference is evident in the IR color, bottom panel). However, due to the age range of the cluster population we are studying, the impact is negligible compared to having self-consistent treatment of the contributing nebular luminosity. More than 50 \% of the clusters in the optical color are located in areas of the diagram that can only be reproduced by models that include the photoionized gas.  

Finally, the inclusion of the IR color (bottom panel), clearly show that clusters with a flux excess at the redder wavelengths are mainly located in an area of the diagram where the $R-I > 0$ and $H-R > 1.0$ mag. The IR color of the clusters with a red excess suggests that the extinction in these objects should be much higher than the one predicted by the optical and UV colors. In other words, at the NIR wavelengths it is possible that we are probing a different, more deeply embedded stellar populations, indicating that these are very young clusters. We will discuss the cluster properties and possible connections with the origin of the red excess in Section \ref{rex-cause}.  

\subsection{Age, mass, and extinction of the cluster population}

\begin{figure}
\resizebox{\hsize}{!}{\rotatebox{0}{\includegraphics{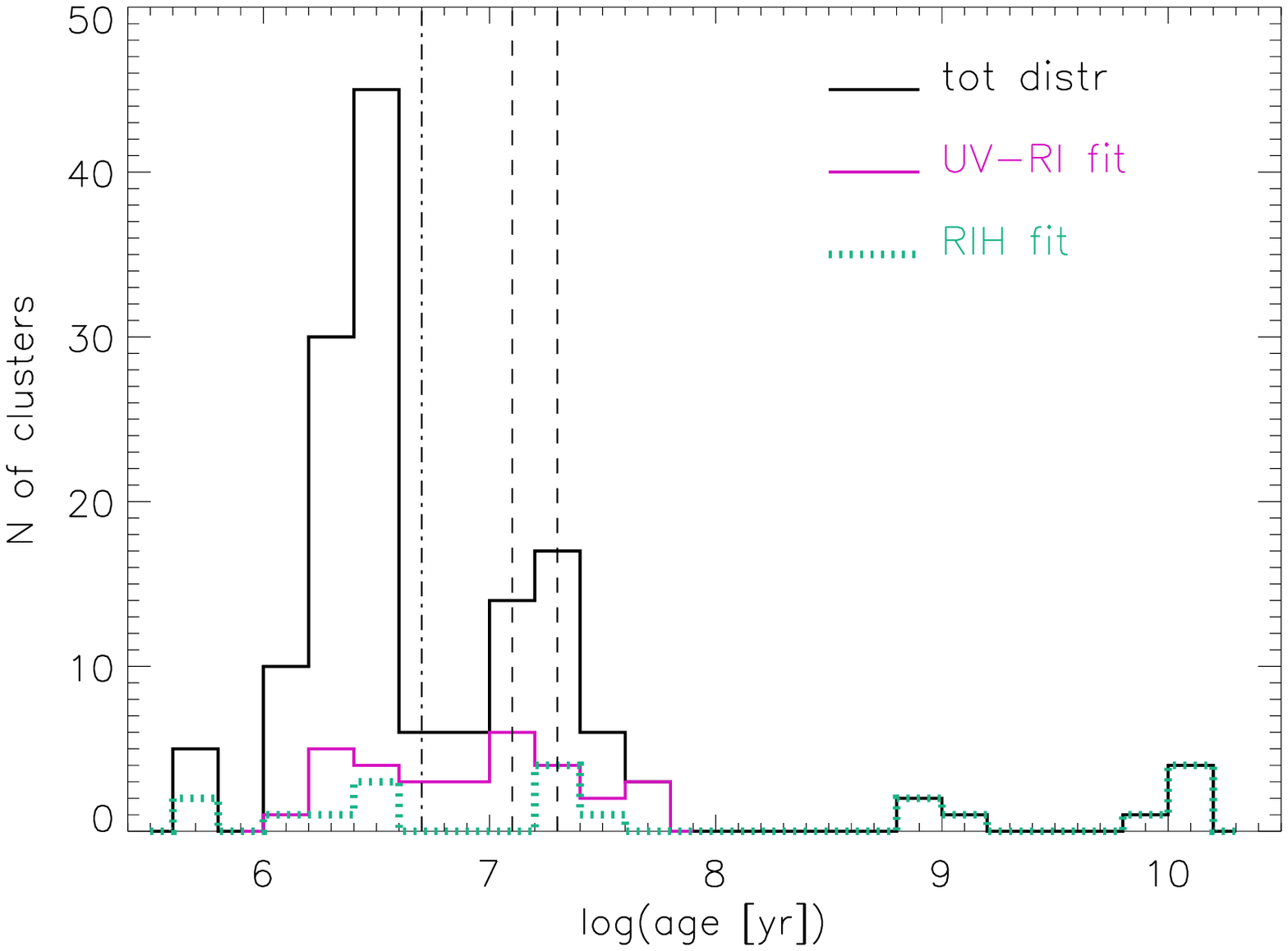}}}\\
\resizebox{\hsize}{!}{\rotatebox{0}{\includegraphics{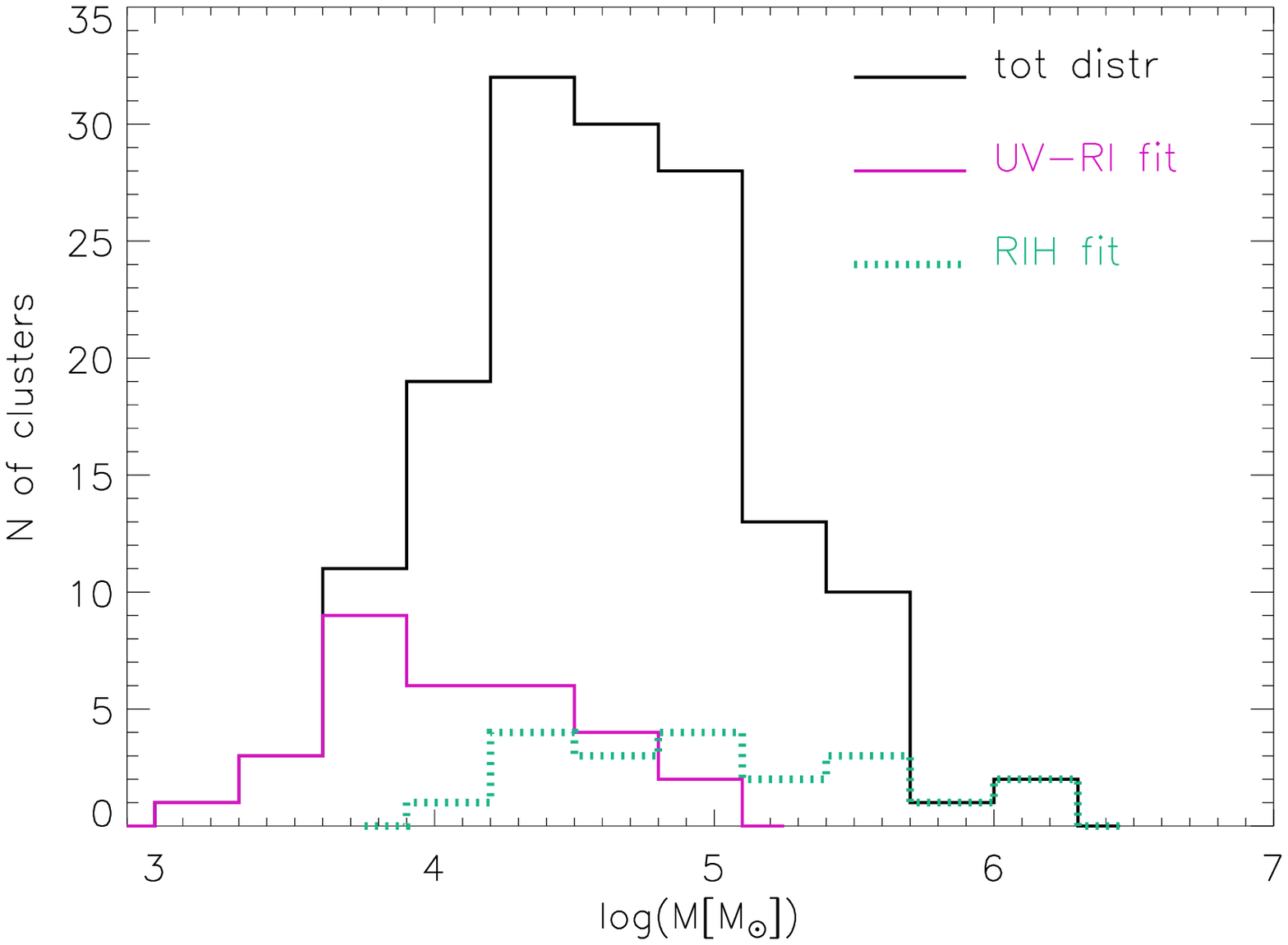}}}\\
\resizebox{\hsize}{!}{\rotatebox{0}{\includegraphics{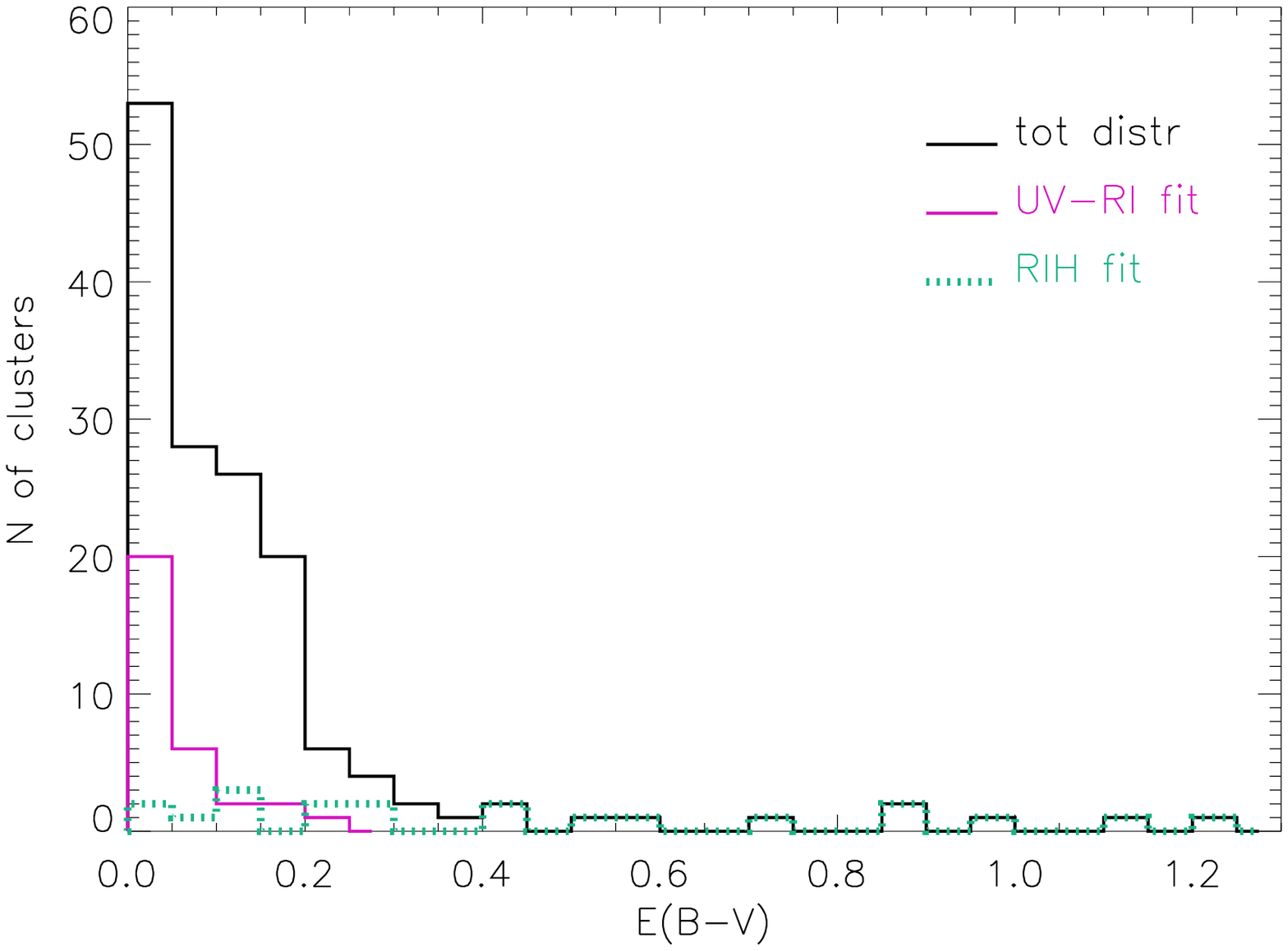}}}\\
\caption{Final age, mass, and extinction distributions for 150 clusters in solid black line histograms. The plots include also the individual distributions of the 20 clusters detected in $R$, $I$, and $H$ bands (dotted thick green line) and the 31 clusters for which properties were found using $FUV$, $R$, and $I$ detections (solid purple line). See the main text for details.}
\label{age-mass-ext}
\end{figure}

In this section we present the derived age, mass, and extinction distributions of the cluster population. For 57 \% of the systems we had at least 4 different data points (integrated fluxes) to perform the SED fit. The remaining fits were obtained from only 3 data points: 13 \% with $R$, $I$, and $H$ fluxes (hereafter, $RIH$ fit); 20 \% with $FUV$, $R$, and $I$ (hereafter, $FUV-RI$); 10 \% with a detection in $B$ or $U$ and $R$, and $I$.

The uncertainties associated with the recovered values depend on the used filter combinations; the photometric errors (including uncertainties produced by the quality of the data and the reduction process); the models (limited by our ability to reproduce the luminosity properties of the clusters, by assumptions on the initial mass function - IMF - and on the mass-to-light ratio); and the applied attenuation law. 

\begin{figure}
\resizebox{\hsize}{!}{\rotatebox{0}{\includegraphics{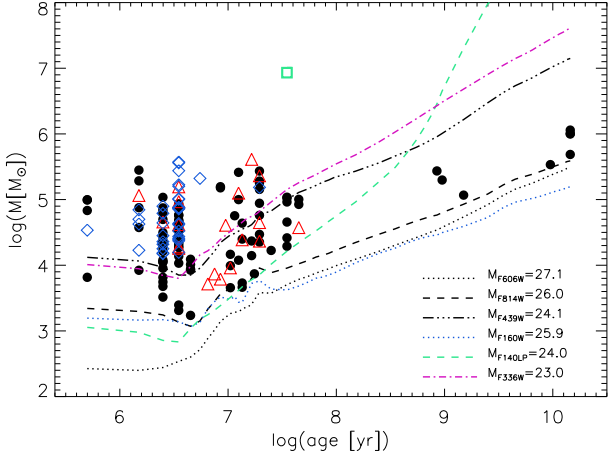}}}\\
\caption{Mass-age diagram. Clusters which do not show any red excess are represented with filled black dots. Clusters affected by IR excess are shown with red triangles. The blue diamonds are for clusters affected by NIR excess. The lines (see the insert) show the detection boundaries in ages and masses corresponding to the magnitude limits reached at $\sigma\leq 0.2$ in each filter. As already expected there are no clusters sitting below these limits. The green square shows the mass and age of the cluster-like object presented in Section \ref{SED_par}.}
\label{age-mass}
\end{figure}

 In the cluster analysis, an important factor of uncertainty  is produced by  stochastic effects in the sampling of the stellar IMF. Several works \citep[e.g.][among many others]{2004A&A...413..145C, 2009ApJ...699.1938M, 2011arXiv1101.4021S} have pointed out that stochasticity becomes important in low mass clusters ($< 10^5$ $\msun$) and may drastically affect the photometric studies of  clusters with masses lower than $10^4 \msun$. The spectral evolutionary models of star clusters are built assuming a continuously populated IMF. In reality, in low mass clusters the stars randomly sample the IMF, meaning that they may have formed  a number of massive stars different from the predicted ones by the models. Massive stars dominate the colors of the clusters. Recently, \cite{2011arXiv1101.4021S} have showed that the luminosity properties of the clusters of $ < 10^4 \msun$ change so drastically, that the uncertainties in the age can be as big as 2.5 dex. A wrong age estimate has important consequences for the  mass as well (e.g. if the age is overestimated, the mass will also be bigger than the actual value because of the increasing mass-to-light value as function of the age).  However, we show in Figure~\ref{age-mass-ext} and \ref{age-mass} that many clusters have masses higher than $10^4 \msun$ suggesting that the stochastic effects, although present, are not very important \citep[the error in the age estimate is about 5-10 \%][]{2004A&A...413..145C}. In Section~\ref{rex-cause}, we again address stochasticity to explain the NIR excess observed in some clusters which luminosity could be dominated by red super giant stars (RSGs). 

 In \citet{A2010c}, we discussed the uncertainties introduced by the set of data on the recovered estimates using Monte Carlo (MC) simulations. Since the data set, the models, and the extinction law applied are the same as the ones used in the analysis performed on  Mrk\,930, we apply those recovered uncertainties to the present results.

The recovered age distribution shows, as expected, a very young cluster population (upper panel in Figure~\ref{age-mass-ext}). The most populated age bin is the 3-4 Myr one ($\log$(age)$=6.5$ Myr), similar to what already observed in Haro 11 \citep{A2010} and ESO 185-IG13 \citep{A2010c}. From MC tests \citep{A2010c} we know that the robustness of this peak is quite high (this age is recovered in 99 \% of the cases). 

A secondary peak is found at slightly older ages, between 10 and 20 Myr (signed with vertical black dashed lines, $\log$(age)$=7.1$ and 7.3 Myr). These two age bins are the most uncertain in our analysis. Independently of the filter combination used, we observed in the MC tests that the recovered number of systems in the 10-20 Myr bins are doubled with respect the initial number of objects, meanwhile the age range between 20 and 150 Myr are emptied. The causes of this behaviour is the loop in the evolutionary tracks at ages between 10 and 40 Myr (see upper right and bottom panels in Figure~\ref{ccd}) and a degeneracy between age and extinction between 40 and 150 Myr, i.e. the extinction vector lies roughly parallel to the evolutionary tracks (see also Section 3.3 in \citealp{A2010c}). Estimates based on a 3-data point SEDs are, in general, even more uncertain. The $FUV-RI$ fit produces two spurious peaks at 12 Myr ($\log$(age)$=7.1$) and at 1 Gyr ($\log$(age)$=9.1$). The age distribution of clusters with a $FUV-RI$ fit (purple solid line histograms, upper panel in Figure~\ref{age-mass-ext}) has a considerable fraction of objects in the 12 Myr bin. Likely, two thirds of these objects are in reality older. Similarly, the $RIH$ fit produces spurious peaks at $\log$(age)$=6.7$ (dash-dot line in Fig. \ref{age-mass-ext}) and $\log$(age)$=7.3$ Myr. We see, however, that the number of objects in these two bins is small. We conclude that likely the secondary peak is not securely established and open for a scenario where cluster production in the galaxy has started roughly 100 Myr ago and continued until the present time.  Two much older peaks at 1 and 14 Gyr prove that the present starburst phase is not the first burst event in the galaxy.

The cluster mass distribution (middle panel in Figure~\ref{age-mass-ext}) covers a wide range of masses, from $10^3 \msun$ to $\sim 10^6 \msun$. We observe that more than 20 \% of clusters are as massive or more massive than $10^5 \msun$ (the super star cluster range).
MC tests showed that a significant fraction (roughly 30 \%) of recovered masses are underestimated. The difference between input and recovered masses is, however,  not higher than a factor of 2 (when $FUV-RI$ and $RIH$ fits are performed). Only in the case of a $RIH$ fit we noticed \citep{A2010c} that in 20 \% of cases the mass was overestimated up to a factor of 2.
  
 The mass-age plot (Figure~\ref{age-mass}) shows the whole cluster population. In contrast with Haro 11 and ESO 185-IG13, we find no young clusters with masses $\sim 10^6 \msun$. Clusters with a NIR excess are all (except one) younger than 6 Myr. IR excess clusters have ages between 1 and 35 Myr. Even in Mrk\,930 we observe a lack of low mass ($< 10^4 \msun$) and very young clusters ($<$ 4 Myr). As already discussed in \citet{A2010, A2010c}, a fraction of these objects is likely not detected due to blending and crowding. We also expect that a considerable number of these clusters are still partially embedded in their birth cocoons of dust and gas. We estimated that a rather moderate extinction of $E(B-V)=0.5$ mag ($A_V \approx 1.5$) is enough to make a cluster of mass below $10^4 \msun$ undetected at visual wavelengths.

\begin{figure}
\resizebox{\hsize}{!}{\rotatebox{0}{\includegraphics{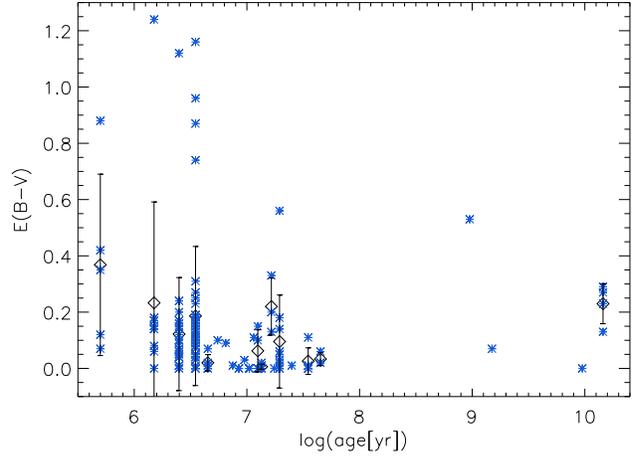}}}\\
\caption{Extinction-age diagram. The blue (grey) asterisks show each single cluster. Superimposed the mean extinction estimated at each model age step and  error bars given by the corresponding standard deviations at each age bin.}
\label{age-ext}
\end{figure}

 The extinction distribution (bottom panel in Figure~\ref{age-mass-ext}) shows, as expected, that more than 80 \% of the clusters have extinction $E(B-V) \leq 0.2$ mag. The remaining fraction is distributed in somewhat higher values, reaching 1.2 mag. Clusters with extinction higher than 0.4 mag could only be analysed with a $RIH$ fit. The extinction-age plot (Figure~\ref{age-ext}) clearly shows that those clusters with higher extinction are, in all cases (except 2), systems younger than 10 Myr. Since these clusters are only observed at the redder wavebands they are likely still partially embedded.  Similar trends in the extinction distribution and extinction-age diagram have also been observed in the other BCGs, Haro 11 and ESO 185, and in systems like the Antennae \citep{2005A&A...443...41M} and M51 \citep{2005A&A...431..905B}.

\subsection{The red excess in young star clusters}
\label{rex-cause}

\begin{figure}
\resizebox{\hsize}{!}{\rotatebox{0}{\includegraphics{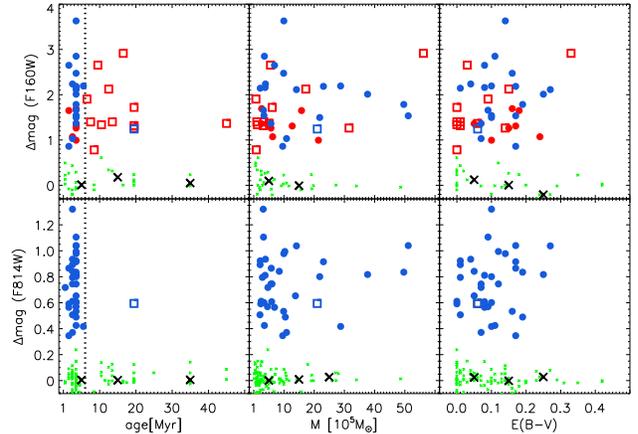}}}
\caption{Strength of the flux excess in $H$ and $I$ bands as function of the cluster properties  derived from fitting the blue side of the SEDs. The vertical dotted lines  in the left side panels separate clusters younger than 6 Myr (filled dots) from older clusters (open squares).  In red we show clusters affected by IR excess ($H$ band), and in blue the ones with NIR excess ($I$ and $H$ bands). The black crosses show, for each bin, the median value of the residuals for normal clusters in the sample, represented by the underlining small green ones.}
\label{delta-exc}
\end{figure}

As already done in the previous studies of Haro 11 and ESO 185-IG13, we try to trace any potential relation between the UV and optical determined cluster properties and the intensity of the flux excess with respect to the best fitting models. In Figure~\ref{delta-exc} we plot the flux excess (residuals) against the corresponding age, mass, and extinction. Represented by dots are the clusters younger than 6 Myr and squares older systems. The color of the symbols are related to the type of excess constrained. If the observed excess is a $\lambda > 1.0$ $\mu$m they are in red. If the excess is longword $\lambda > 0.8$ $\mu$m they are in blue. As comparison, we include the distribution of the "normal" clusters, that are objects with regular SEDs (represented by the black and green crosses).

If we look at the age, mass, and extinction of the clusters versus their residuals ($\Delta m = m_{\textnormal{mod}}-m_{\textnormal{obs}}$) in the $H$ band (upper panels), we do not see any clear relation between the UV-optical properties and the strength of the IR excess. We already noticed from the IR-optical color-color diagram (bottom panel, Figure~\ref{ccd}) that the clusters affected by a red excess require much higher extinction values than the ones from the UV-optical colors (see the other two diagrams) to move toward  the areas of the tracks corresponding to the estimated age. This property suggests that another mechanism contributes at the IR waveband and not in the optical.

The same analysis, but including the $I$ band residuals shows a clear correlation between the age of the cluster and the $I$ band excess (bottom panels in Figure~\ref{delta-exc}). In all except one case, the $I$ band is found in clusters younger than 6 Myr. The spread in mass and extinction is, however, similar to the trend observed in the upper panels.  

The age of the clusters (even if determined by UV and optical luminosities) may give insights on the origin of the red excess. Possible origins for the NIR excess have been widely discussed in \citet{A2010, A2010c}. Here we will shortly summarize the main aspects.

\subsubsection{The $I$ band excess}

Due to the narrow cluster age range where the $I$ band excess is found, a viable explanation for this feature is the extended red emission (ERE, see for a review \citealp{2004ASPC..309..115W}).  The ERE is observed as a soft rising continuum between $0.6-0.9$ $\mu$m. It is observed around  galactic and extragalactic H{\sc ii} regions (among many others, \citealp{1995A&A...304L..21P}; \citealp{2000ApJ...544..859G}) and caused by a photoluminescence reaction on dust grains heated by hard UV radiation. Such energetic photons are mainly produced in short-lived massive stars. This could explain why the $I$ band excess in our clusters is over after 6 Myr. A possible contribution from ERE to the $I$ band flux of unresolved clusters was also suggested by Reines et al. (2008b).

\subsubsection{The IR excess}

Several mechanisms can concur to make the flux at $\lambda > 1.0$ $\mu$m higher than predicted by models, which include stellar continuum and nebular contributions. The distance of the galaxy and the resolution achieved - the best with the current accessible facilities - limit our studies to the integrated properties of these massive star clusters. However, observations of much closer resolved clusters and numerical predictions of stellar populations in clusters can give us an hint of the mechanisms which are likely producing the observed red excess.

Among the youngest and massive resolved star clusters, 30 Doradus (hereafter 30 Dor) in the Large Magellanic Cloud (LMC), represents the best reference to understand what a recently born star cluster looks like. 30 Dor is the central region of the extended Tarantula nebula. Multiwavelength studies of this regions have dissected the different components of the complex 30 Dor environment. \citet{1997ApJS..112..457W} identified five different stellar populations in the region: the bright core early O-type stars which are part of the compact star cluster R136; in the north and west region embedded massive YSOs; 3 more evolved stellar population groups in the southern and 1.0' away in the western region. The R136 cluster has a mass of $\sim 10^5 \msun$ and is 3 Myr old  (\citealp{2010ARA&A..48..431P} and references therein). This nuclear region ($\leq 3$ pc) is dust and gas-free. Recent studies \citep{2010MNRAS.408..731C} report that stars more massive than the usual assumed theoretical limit of 120 $\msun$ reside in the core of R136, probing the extreme nature of the star formation process in very massive star clusters. However, the aperture radius we are using in our analysis is much larger (radius of $\approx 36$ pc) than the size of R136. Our apertures are comparable to the size of the image of 30 Dor showed in Figure 1 of \citet{2010MNRAS.405..421C}. This suggests that while the optical range is dominated by the stellar and gas emission, the IR transmits also flux from the diffuse dust, heated by the hard UV radiation field, the embedded YSOs formed in triggered star formation events at the edge of the nucleus, where most of the dense gas and dust filaments are located, and the low mass stars still in the PMS phase (see \citealp{2002AJ....124.1601W}; \citealp{2001AJ....122..858B}). 

In the literature, studied cases of IR excess in young embedded or partially embedded unresolved extragalactic clusters have explained the red excess as due to an important contribution by YSOs \citep{2009MNRAS.392L..16F}, or hot dust (\citealp{2005ApJ...631..252C}; \citealp{2005A&A...433..447C}). Likely, the same mechanism is causing the red excess in young star clusters in BCGs.

After several Myr this complex phase is over, so it cannot explain why we  still observe objects with an IR excess at older ages. For these evolved clusters a possible source of excess can be an important contribution from RSGs. These are rather massive ($\sim 25-7 \msun$) stars. They contribute mainly to the NIR for clusters age between 8 and 60 Myr. Models usually predict the number of RSGs, assuming that the stars in a cluster fully populate the underlying IMF. However, this assumption is not valid, if the cluster is less massive than $10^6 \msun$ \citep{2010ASPC..425...55L}, and causes important variations for cluster masses below $10^5 \msun$ \citep{2004A&A...413..145C}. Moreover, it has been observed that in metal-poor environments the number of observed RSGs tends to be higher than the predicted one from the ratio of blue versus RSGs (see \citealp{2002A&A...386..576E} and \citealp{2001A&A...373..555M}). Therefore, both effects would be observed as a rise in the NIR integrated flux of an unresolved cluster, which our current models cannot fully account for. It is not clear, however, why we do not see any mass dependence between the excess in $H$ band and the mass of the cluster, which could support the stochasticity scenario, or why our models cannot predict the correct number of RSGs only for some of the clusters. NIR spectroscopy is needed to prove direct evidences for these scenarios.

The environment of Mrk\,930 is quite crowded and we cannot exclude that in some cases the red excess could be caused by a newly formed cluster in the same aperture or at the same line of side, visible only in the IR.

Other "exotic" explanations for the excess in clusters could be a bottom heavy IMF, i.e. a much steeper slope at low mass ranges (higher number of low mass stars), or a second stellar population formed after a long time-delay ($\sim 10$ Myr). However, studies of IMF variations in massive star clusters (see \citealp{2010ARA&A..48..339B}; \citealp{2010ApJ...710.1746G}) do not find any sign of variation; and second populations with delay of few tens of Myr have not been observed yet. 

\begin{figure*}
\resizebox{0.48\hsize}{!}{\rotatebox{0}{\includegraphics{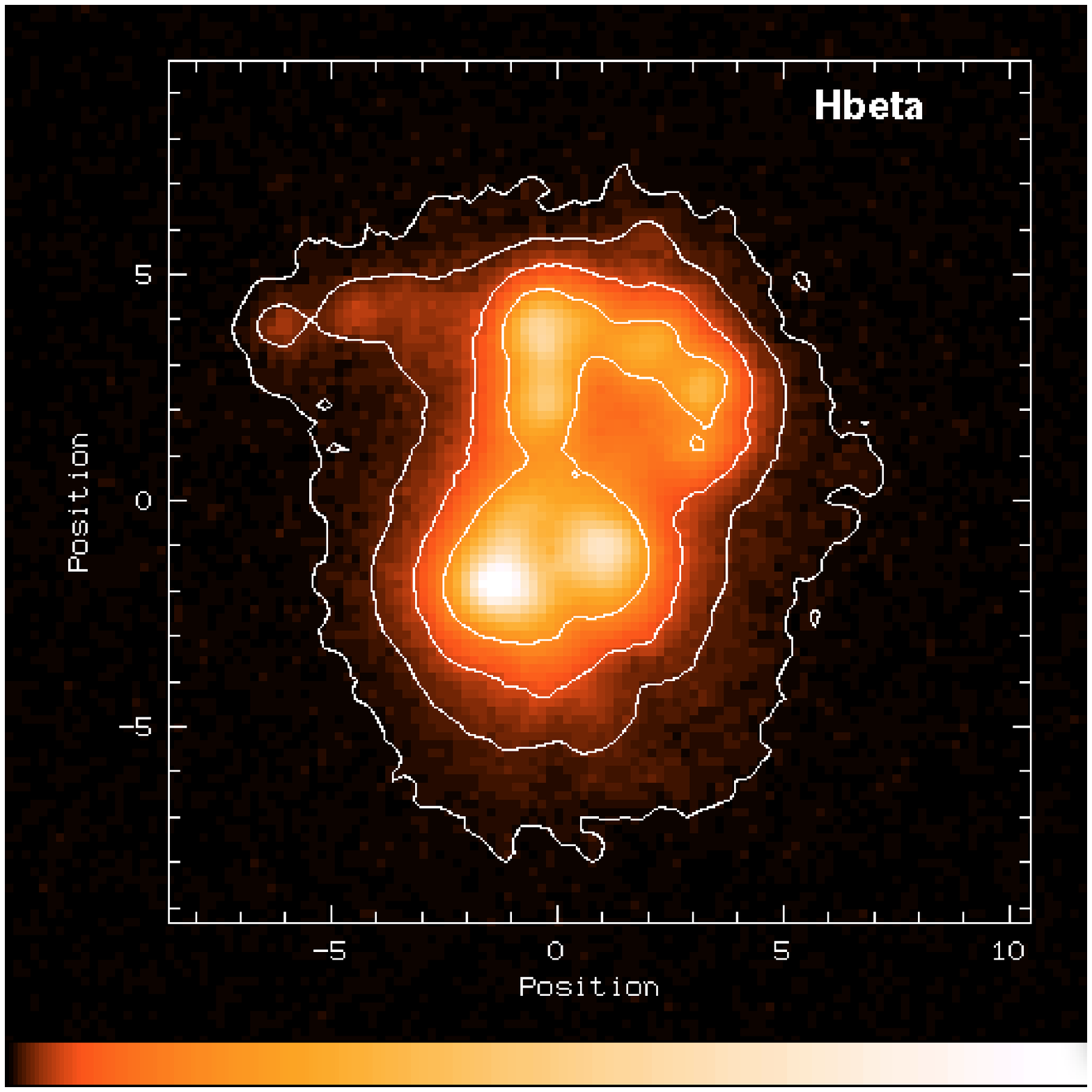}}}
\resizebox{0.48\hsize}{!}{\rotatebox{0}{\includegraphics{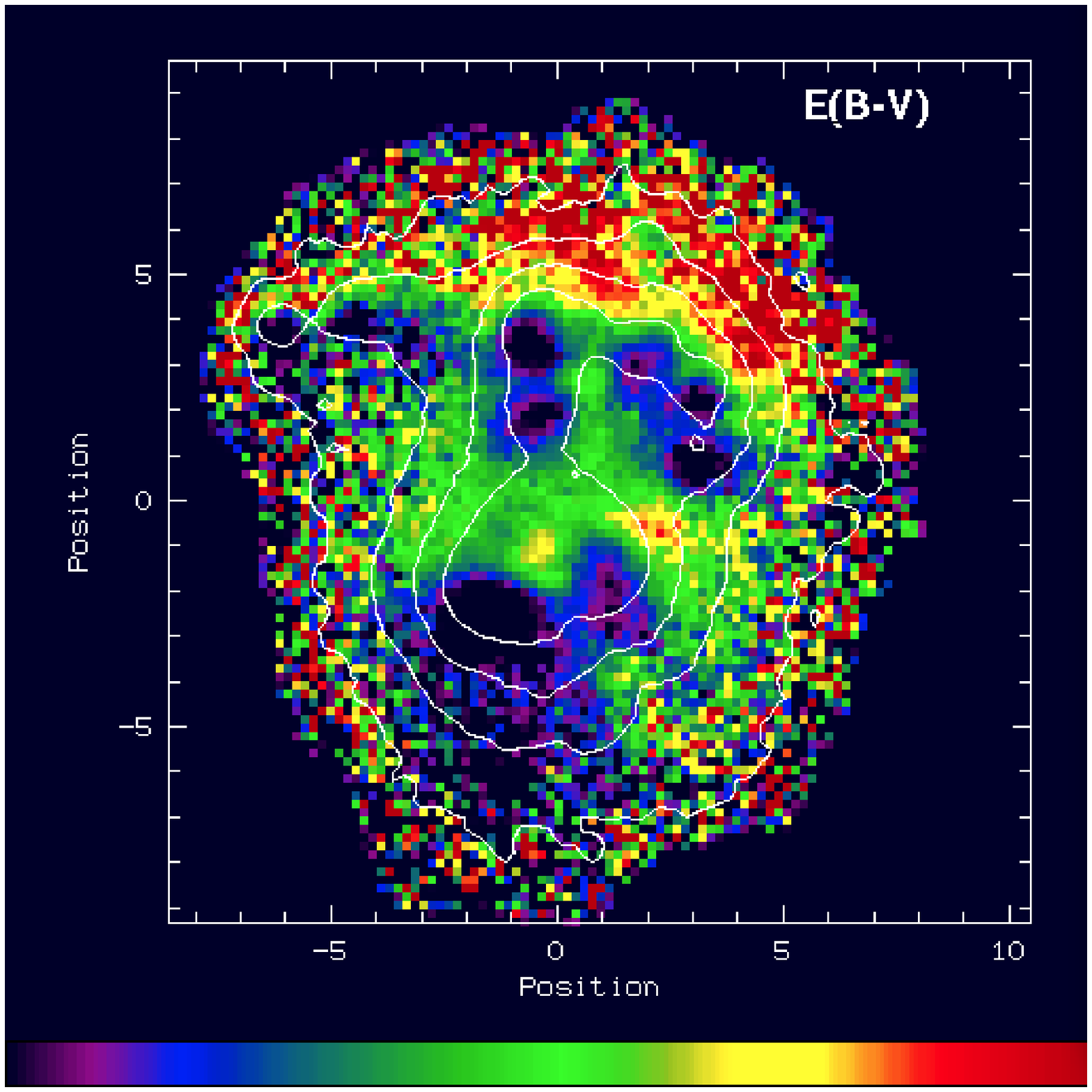}}}
\caption{Left: The \hb \, image of the galaxy in logaritmic scales (from 0 to $5\times 10^{-16}$ erg/s/cm$^2$).  Overplotted in white, the contours at: $10^{-16}$, $3 \times 10^{-17}$,$10^{-17}$,$3 \times 10^{-18}$ erg/s/cm$^2$ flux levels. The coordinate box units are in arcsec. The color bar is in logarithmic scales and goes from f(\hb) value of 0 (black) to 5$\times 10^{-16}$ erg/s/cm$^2$ (white). Right: The extinction map of the galaxy, E(B-V). It is in linear scales and the bar cover from 0 (black) to 0.85 (red), the mapped values. The same \hb \, contours are overplotted.}
\label{ext-map}
\end{figure*}

\section{Mapping the extinction across the galaxy}
\label{ciao}

We have estimated the extinction in the galaxy using three independent analyses: cluster SED fitting,  \ha/\hb \ line ratio from long-slit low-resolution spectroscopy, and narrow band \ha/\hb \ imaging ratio. 
\begin{table*}
  \caption{Measured f(\ha) and f(\hb) in: the galaxy (first raw) using imaging data, and in the 3 main knots using the extracted spectra.  The extinction values, $E(B-V)$, are obtained from \ha \ over \hb \ ratio, assuming a Galactic extinction law. The total \ha \ luminosity, L(\ha) has been estimated assuming a distance of  72.0 Mpc.}
\centering
  \begin{tabular}{|c|c|c|c|c|c|}
  \hline
   &  & f(H$\alpha$) & f(H$\beta$) & E(B-V) & L(\ha)\\
   &  & ($10^{-12}$ erg/s/cm$^2$) &($10^{-12}$ erg/s/cm$^2$) & mag & ($10^{41}$ erg/s)\\
  \hline
Mrk\,930 & imaging& 1.03 &  &  & 7.46\\
Knot A  & spectr & 0.32 $\pm$ 0.08& 0.10 $\pm$ 0.17&0.10 $\pm$ 0.04& - \\
Region B  & spectr & 0.23 $\pm$ 0.07& 0.07 $\pm$ 0.08&0.16 $\pm$ 0.03& - \\
Region C& spectr & 0.35 $\pm$ 0.05& 0.12 $\pm$ 0.09&0.07 $\pm$ 0.03& - \\
  \hline
\end{tabular}
\label{table-ext}
\end{table*}

\begin{figure}
\includegraphics[width=0.5\textwidth, trim=14mm 0mm 10mm 0mm]{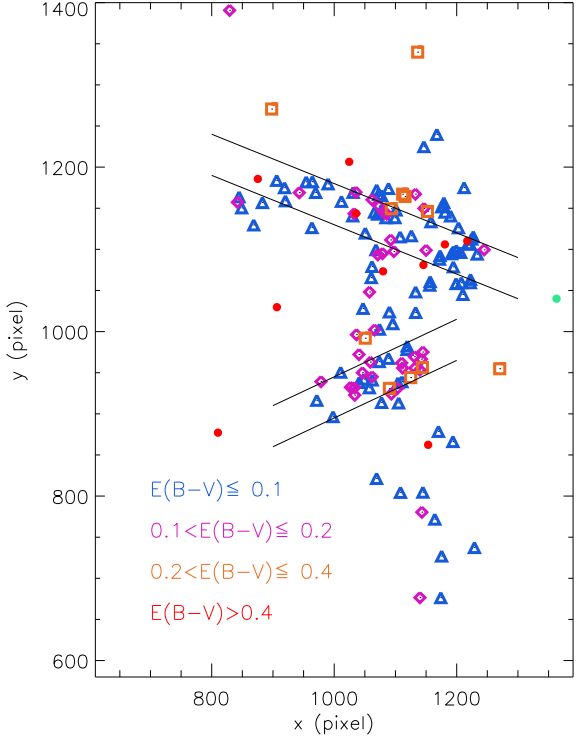}
\caption{The recovered extinction versus the position of the cluster in the galaxy. The inset show the color coding used in the plot and corresponding to each interval of extinction values. The position of the two slits and the regions are indicated. In green dots we represented the excluded cluster-like object from our analysis.}
\label{ext-pos}
\end{figure}

Regions B and C are resolved in the 2-dimensional spectrum (Pos-1), hence we were able to extract a single spectrum for each of them. The knot A was blended in a single blob (Pos-2) so we extracted a single spectrum enclosing the total flux. The fluxes of  the \ha\, and \hb \ lines were measured in the three one-dimensional spectra. To correct the two emission lines from the underlying stellar absorption, we performed a fit to the continuum using the {\sl Starlight} fitting procedure (see Section \ref{ciao4}).  The \ha \ flux was also corrected for the contribution of the blended [N\,{\sc ii}] line at 6548.0 \AA. The flux of this last emission line was estimated from the measured flux of the unblended [N\,{\sc ii}] at 6584.0 \AA \ and the expected line ratio [N\,{\sc ii}]$_{6548.0}/$[N\,{\sc ii}]$_{6584.0}=0.33$ for case B recombination 
\citep{2006agna.book.....O}. The final extinction was estimated using the galactic extinction law \citep{1989ApJ...345..245C}. The line fluxes and the estimated extinction are all listed in Table \ref{table-ext}.

We constructed an $E(B-V)$ map from the H$\alpha$/H$\beta$ images assuming a
non reddened H$\alpha$/H$\beta$ ratio of $2.87$ and the galactic extinction law.
No correction for Balmer absorption was applied, and hence the  $E(B-V)$
is only indicative and may overestimate the extinction in regions where the
H$\beta$ emission  is weak. In Figure~\ref{ext-map} we show from left to right: H$\beta$ image (with logarithmic intensity scaling), and $E(B-V)$ map.

In Figure~\ref{ext-pos}, we have reconstructed the position of the clusters in the galaxy and their recovered extinction. Clusters with $E(B-V) \leq 0.2$ (blue triangles and purple diamonds) are the majority. Clusters with higher extinction are fewer and located around almost dust-free companions. The smaller number of high extinguished clusters could be caused by a bias in the selection method. Since optical filters have been used to create a position catalogue (see Section \ref{ciao3}), it is likely that those redder clusters have been missed. The positions of the two slits are indicated  in Figure~\ref{ext-pos}. The slit is 1" width, however, taking into account the seeing, it is likely that the regions close to the edges of the slit have also contributed to the transmitted spectrum. In general, we observe that the values recovered from the two hydrogen line ratios are very low and suggest that the spectra are dominate by an averaged value of the extinction observed in the clusters.

Similarly, we notice that the resolution of the map in Figure~\ref{ext-map} is much lower than the one reached with the HST data, so the resulting $E(B-V)$ values are averaged over a wider region than a typical cluster size. For this reason we look at general trends and not to one-by-one analogies between a cluster and a pixel-map. Morphologically, we observe a correspondence among the brightest \hb \, regions (left panel, Figure~\ref{ext-map}), the less extinguished areas of the galaxy (right panel), and the location of the very young and, in most cases low-extinguished, star clusters (see also Figure~\ref{pos-gal}). These areas are also the ones which dominate the signal of the spectra, because of the very low $E(B-V)$ values obtained from the optical line ratios. This feature suggests that feedback from young star clusters has likely cleaned (destroyed) the dust in these regions (a large scale of the 30 Dor core). We observe a gradual increase of the extinction around the dust-free knots. The centre (between the two bright clumps) and the most northern regions of the galaxy are the most extinguished. We only detect a dozen of clusters in these two regions:  many of them have very low extinction, or not bigger than 0.4 mag, suggesting that highly extinguished clusters in this area have not been detected. 

We conclude that the extinction maps of the galaxy agree fairly well on the average values, despite having been produced with different techniques. Locally, however, only the resolution reached by the study of the clusters allow us to explore extinction variations on much smaller and detailed scales.
Of course, there is no a priori reason why the extinctions derived from nebular lines and continuum should be the same even if we could match the HST resolution since these spectral components may have different contributions from regions along the line of sight.

\section{Probing the starburst properties and the formation history in Mrk \, 930}
\label{ciao2}

\subsection{The bursting nature of Mrk\,930}

To probe the bursting nature of Mrk\,930 we constrain the $b$-parameter and the gas consumption time scale, both considered starburst tracers.

The visual magnitude of Mrk\,930 is M$_V=15.12$ mag (Micheva et al. in prep), which corresponds to a total luminosity of L$_V=3.89\times10^9$ L$_{\odot}$. A more detailed investigation of the underlying stellar population will be presented by Micheva et al (2011 in prep)
but to get an approximate estimate of the total stellar mass in Mrk\,930 we take a mass-to-light ratio, M/L$=0.87$ which is the median
for the five luminous BCGs with well constrained stellar masses in \cite{2001A&A...374..800O}. We estimate in this way the total stellar mass in the galaxy, $M_{\star}\sim3.4\times10^9$ $\msun$. \citet{2002AJ....124..862H} estimated the SFR in Mrk\,930 using the luminosity of the galaxy at 60.0 $\mu$m (9.1$\msun$/yr) and at 1.4 GHz (10.6 $\msun$/yr). Estimates of the SFR using the \ha\, luminosity give lower values, i.e., SFR(\ha)$=5.34 \pm1.79$ $\msun$/yr \citep{2009MNRAS.399..487R}. This value is in very good agreement with the estimate we obtained from the measured L(\ha) shown in Table~\ref{table-ext}. Applying the Kennicutt relation \citep{1998ARA&A..36..189K}, we determine a slightly higher SFR(\ha)$=5.89$ $\msun$/yr. 

We constrain an upper limit of the so-called $b$-parameter, i.e. the ratio of the current SFR to the average SFR of the past in the galaxy \citep{1986FCPh...11....1S}. Galaxies with constant or declining SFRs have $b \leq 1$, while systems which are undergoing a violent burst episode have $b \gg 1 $ \citep{1994ApJ...435...22K}. Using the estimates of the $M_{\star}$ and of the SFR(\ha), we find $b \approx 16$ for Mrk\,930, confirming that the this galaxy is truly a starburst. A similar conclusion can be reached from the gas consumption time scale, defined as the ratio between the available gas and the current SFR in the galaxy, i.e. $\tau_{gas}=$M$_{gas}/$SFR.  \citet{2002AJ....124..862H} estimated a mass of neutral hydrogen of M$_{HI}=3.02\times10^9$ $\msun$. There are not available estimates of the molecular and ionized gas mass in the galaxy, so we assume that the total amount of gas is 2 times the determined M$_{HI}$ \citep{2005ApJ...625..763L}. We find then that Mrk\,930will be able to sustain the current SFR for approximately another half Gyr ($\tau_{gas}=1$ Gyr).

\subsection{Starburst propagation revealed by the star cluster ages}

To trace the starburst properties and the formation history of Mrk\,930, we compare the cluster properties with other probes: the [O\,{\sc iii}]/\hb \, ratio map of the galaxy and the fit to the stellar continuum spectra of the 3 starburst knots (next sections).

\begin{figure}
\includegraphics[width=0.5\textwidth, trim=14mm 0mm 10mm 0mm]{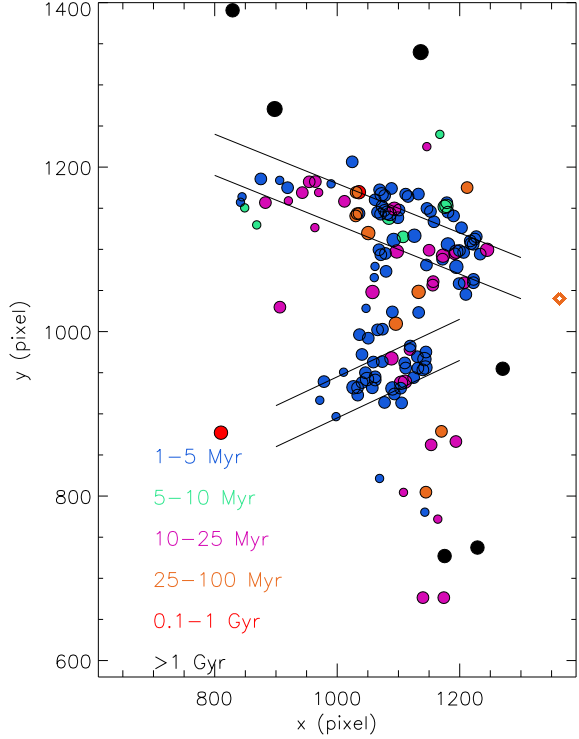}
\caption{Position of the clusters in the galaxy as function of their ages (see color coding in the inset). The size of the dots is proportional to the estimated mass, i.e., bigger radii are for higher masses. The position of the two slits and the regions are indicated. The age range between 10-25 Myr (shown in purple) is the most uncertain and a fraction of these clusters is likely older ($<100$ Myr). We also notice that the clusters in the age range 25-100 Myr (orange dots) are, in reality, all younger than 50 Myr (see Figure~\ref{age-mass-ext}). In red, we show the few objects detected at roughly 1 Gyr and in black, the ones at 14 Gyr. See main text for details.}
\label{pos-gal}
\end{figure}

\begin{figure}
\resizebox{\hsize}{!}{\rotatebox{0}{\includegraphics{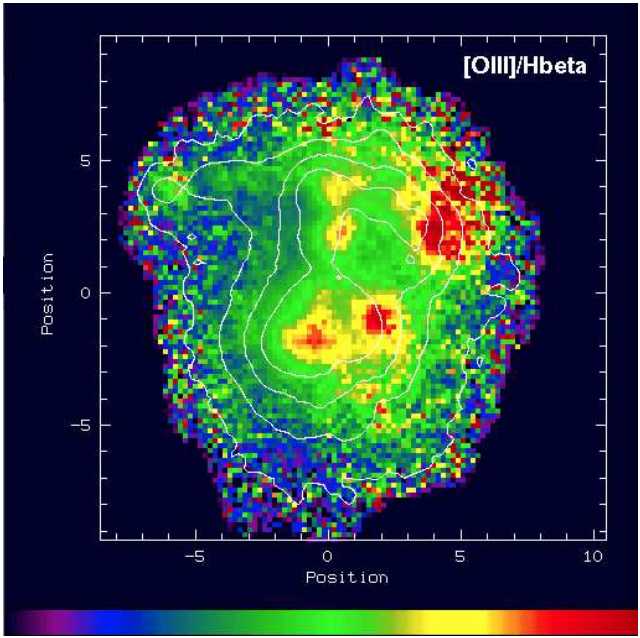}}}
\caption{The [O\,{\sc iii}]/\hb \, map of the galaxy produced with narrow band imagings.  Overplotted in white the \hb contours at: $10^{-16}$, $3 \times 10^{-17}$,$10^{-17}$,$3 \times 10^{-18}$ erg/s/cm$^2$. The coordinate box units are in arcsec. The color bar is linear and goes from [O\,{\sc iii}]/\hb \, value of 0 (black) to 10 (red). See main text for details.}
\label{oiii}
\end{figure}

Using the age of the clusters we mapped the cluster formation and propagation in the galaxy (see Figure~\ref{pos-gal}). We clearly observe that the starburst regions are dominated by very young star clusters (blue dots) with an age $< 5 $ Myr. The [O\,{\sc iii}]/\hb \, ratio is known as a tracer of the mean ionisation level and temperature of the photoionized gas \citep{1981PASP...93....5B}. The stronger the UV radiation field is and the higher the  [O\,{\sc iii}]/\hb \, ratio becomes. In Figure~\ref{oiii} one can easily observe that the areas with higher ratio values are the ones where the younger star clusters are found. \citet{1998ApJ...500..188I} estimated for Mrk\,930 a  [O\,{\sc iii}]/\hb \, ratio of $\sim 4.5$, in good agreement with an average estimate from our map and the classification of this target as a H{\sc ii}  galaxy.


We observe that in the burst regions, very young star clusters are located around central older clusters, suggesting a progressive propagation of the cluster formation.  In the extended tail, south of knot A, we mainly find older clusters in agreement with the color shown in  Figure~\ref{h11b}. The tidal tail and region C are separated by a group of older clusters all  located in a small area. The cluster population in the tail is rather young and spans an age range of 1-25 Myr.

\begin{figure*}
\resizebox{0.48\hsize}{!}{\rotatebox{0}{\includegraphics{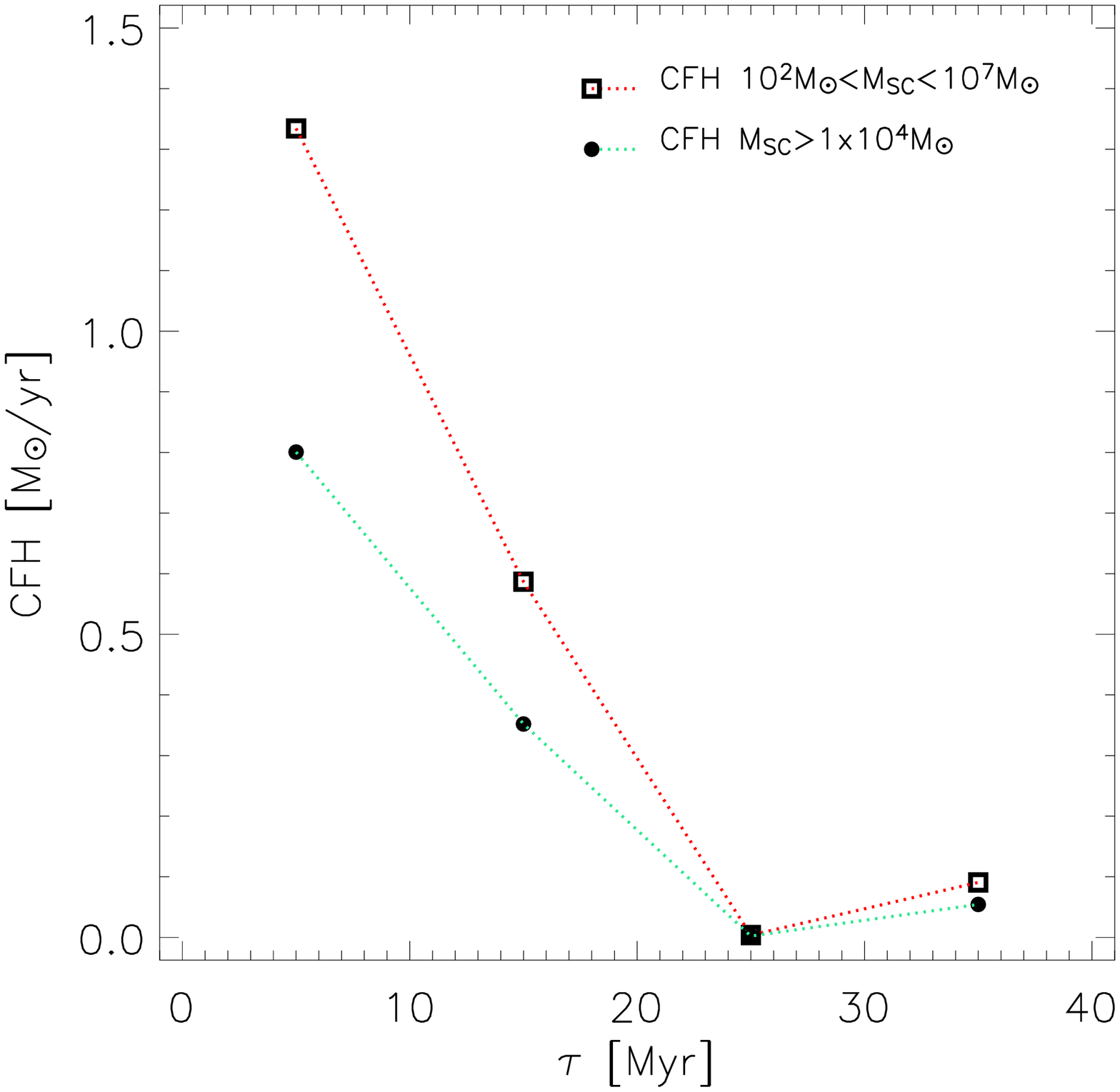}}}
\resizebox{0.48\hsize}{!}{\rotatebox{0}{\includegraphics{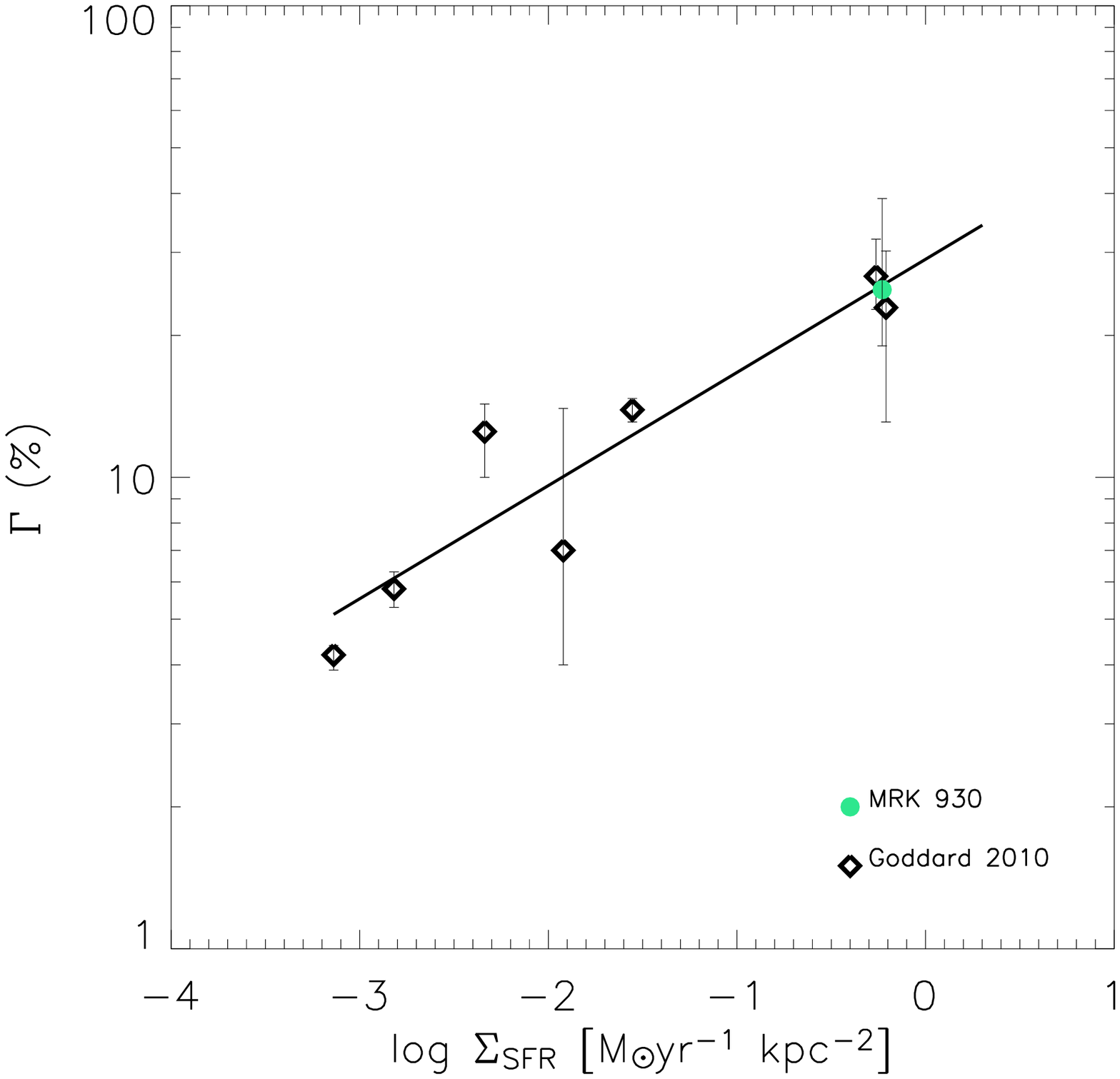}}}
\caption{Left panel: Cluster formation rates during the last 40 Myr of starburst activity. The filled dots connected by the green dotted line show the observed CFR derived from clusters more massive than $10^4 \msun$. The thick squares show the derived CFR, if the total mass in clusters less massive than $10^4 \msun$ is extrapolated using a CMF with index $-2.0$ down to $10^2 \msun$. Right panel: The cluster formation efficiency ($\Gamma$) versus the star formation rate surface density, $\Sigma_\mathrm{SFR}$, using data from \citet{2010MNRAS.405..857G} (black diamonds and the dashed line). The  positions of Mrk\,930 is shown by the green (grey) dot. See text for details.}
\label{CFR-plot}
\end{figure*}

Assuming that we are complete in detecting clusters more massive than $1\times 10^4 \msun$, we derive the CFH during the last 40 Myr of star formation in the galaxy. We consider that a full population of clusters forms with a CMF power-law of index $2.0$ and extrapolate the missing fraction of cluster mass (see \citealp{A2010} for details). We obtain then a trend of the cluster formation rate in the last 40 Myr (left panel in Figure~\ref{CFR-plot}), which shows a clear increase in the rate of cluster formation during the last 20 Myr. The present CFR is $\sim1.3 \msun/$yr, i.e., 1.3 $\msun/$yr of stars are formed in bound star clusters. The previous significant burst was between 30 and 40 Myr ago.

Comparing the present CFR to the SFR(\ha), we find that $\sim$ 25$\pm$10 \% of the SFR is happening in bound clusters ($\Gamma =$CFR/SFR$= 0.25$, see \citealp{2008MNRAS.390..759B}). The fraction is quite high, but comparable with the values we have found for Haro 11 and ESO 185-IG13. \citet{2010MNRAS.405..857G} presented an empirical relation between the cluster formation efficiency, $\Gamma$, and the surface density of SFR, $\Sigma_\mathrm{SFR}$ in the host. This relation suggests that the SFR and the rate of the cluster formation are positively correlated, or in other words, that an enhancement in the SFR increases the number (mass) of the produced clusters.

We estimated a radius of the starburst area of the galaxy integrating the luminosity flux in $B$ band within increasing aperture radii, up to include the outskirts of the galaxy. We observed that the radius, where 80 \% of the light produced in the galaxy is collected, included the whole starburst area. We used this radius, $r_{80\%}\approx1.7$ kpc, to estimate the $\Sigma_\mathrm{SFR}$ in Mrk\,930 and found $\Sigma_\mathrm{SFR}=0.55 \msun$/yr.

In the right panel of Figure~\ref{CFR-plot}, we show the relation obtained by \citet{2010MNRAS.405..857G} and the position of Mrk\,930. The agreement is good, showing evidence of the a positive correlation between the mean properties of the host environment and the production of star clusters. The same correlation has been found in the others BCGs \citep{A2010, A2010c}, probing that the BCG environment, perturbed by the recent merger event, forms a higher fraction of stars in bound clusters. The results obtained by \citet{2010arXiv1009.1618W}, by means of numerical simulations, support this scenario. They show that regular patterns, like spiral arms and shear, tend to fragment the collapsing GMCs and favour the formation of less clustered systems (OB associations). On the other hand, in a merging system, the external pressure on the GMC yields a faster  collapse and enhance the local star formation efficiency inside the GMC.

Finally, we notice that in our analyses (Haro 11, ESO 185-IG13, Mrk\,930) has not been possible to determine which fraction of the clusters we have studied are real bound clusters following the definition by \citet{2010MNRAS.tmpL.168G}. The Goddard et al. relation has been obtained using similar analyses. We cannot exclude that an important contamination by systems which are gravitationally unbound could create this apparent relation. Unfortunately, with the current data we are not available to test whether the relation would still hold or disappear, hence suggesting a universal cluster formation efficiency \citep{2008MNRAS.390..759B}. 

We will discuss this point and the starburst properties of the BCGs in a forthcoming paper (Adamo et al. in prep.).

While the current cluster formation is very high in Mrk\,930, we find a small number of clusters at age  $\sim 1$ Gyr and at $\sim 10$ Gyr. In the age range around 1 Gyr we find a total cluster mass $5.4\times10^5 \msun$ (clusters more massive than $10^5 \msun $). The expected mass in clusters below $10^5$ is negligible, assuming that clusters in this age range has a Gaussian CMF ($M_{\rm peak}=2\times 10^5 \msun$, $\sigma = 0.6 \log(M) \msun$) . 
For  old (age $\sim10$Gyr)  GCs we see a total mass of $4.5\times 10^6 \msun$ when including clusters with masses down to  $3\times10^5\msun$. Assuming the same Gaussian CMF as in the previous example, the total estimated mass for the
old GC population is $1.7\times10^7\msun$. The stellar mass of the galaxy is $M_{\star}\sim3\times10^9$ $\msun$ which would imply that the GCs in Mrk\,930  make up $\sim 1$ \% of the total host stellar mass. This value is smaller than what has been found for ESO\,338  \citep{2003A&A...408..887O} and ESO 185 \cite{A2010c}.
 
\subsection{{\sl Starlight} fit to the starburst knot spectra: the SFH}
\label{ciao4}

Star clusters are produced in the same global star formation event which enriches the host of several stellar populations. Using the {\sl Starlight} population synthesis code\footnote{STARLIGHT \& SEAGal: http://www.starlight.ufsc.br/} 
(\citealp{Fernandes05}; \citealp{Fernandes07}; \citealp{Mateus06}, \citealp{2007MNRAS.381..263A}; see \citealp{2009ApJ...707.1676C} for an applications of {\sl Starlight} to BCGs) 
on our spectra for regions A through C, we try to shed further light on the star formation history (SFH) 
of Mrk\ 930 and check for a consistency the age distribution of
stellar clusters and discrete episodes of enhanced SFR, as revealed from
spectral synthesis.

{\sl Starlight} synthesizes the observed SED
of the stellar continuum as due to a linear combination of single stellar
populations (SSPs) of different age and metallicity.
The SSP library used is based on stellar models by \citet{2003MNRAS.344.1000B} for a Salpeter IMF, three stellar metallicities 
($Z_{\odot}$/19, $Z_{\odot}$/4.75 and $Z_{\odot}$/2.4, assuming $Z_{\odot}$=0.019)
for 59 ages (between 0.25 Myr and 13 Gyr). 

\begin{figure*}
\resizebox{0.8\hsize}{!}{\rotatebox{0}{\includegraphics{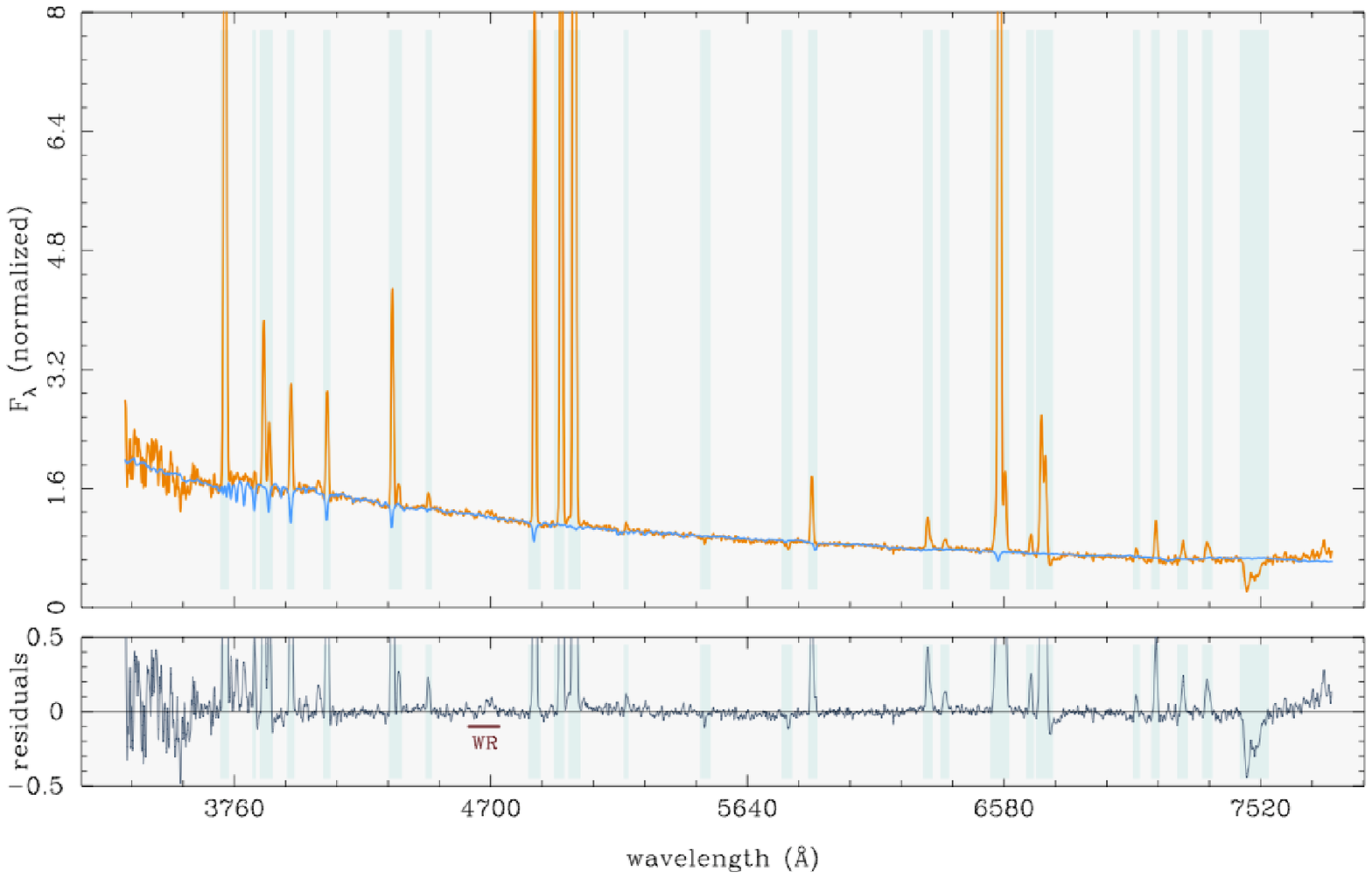}}}\\
\resizebox{0.33\hsize}{!}{\rotatebox{0}{\includegraphics{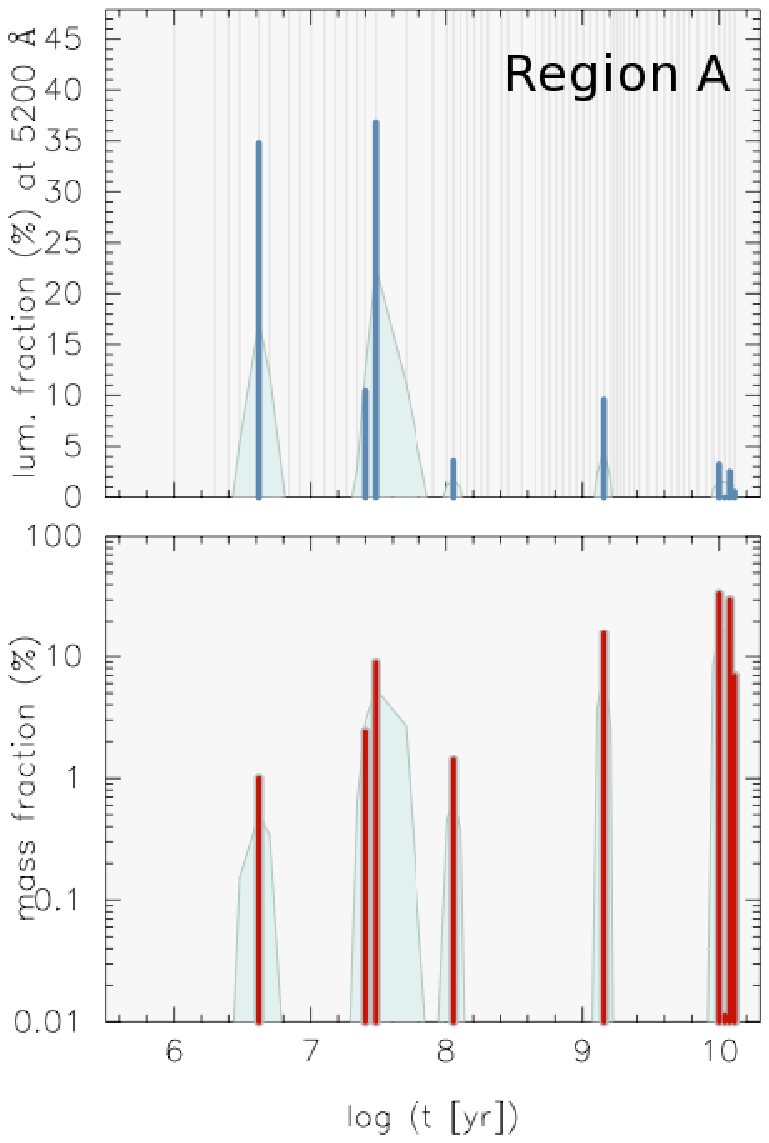}}}
\resizebox{0.33\hsize}{!}{\rotatebox{0}{\includegraphics{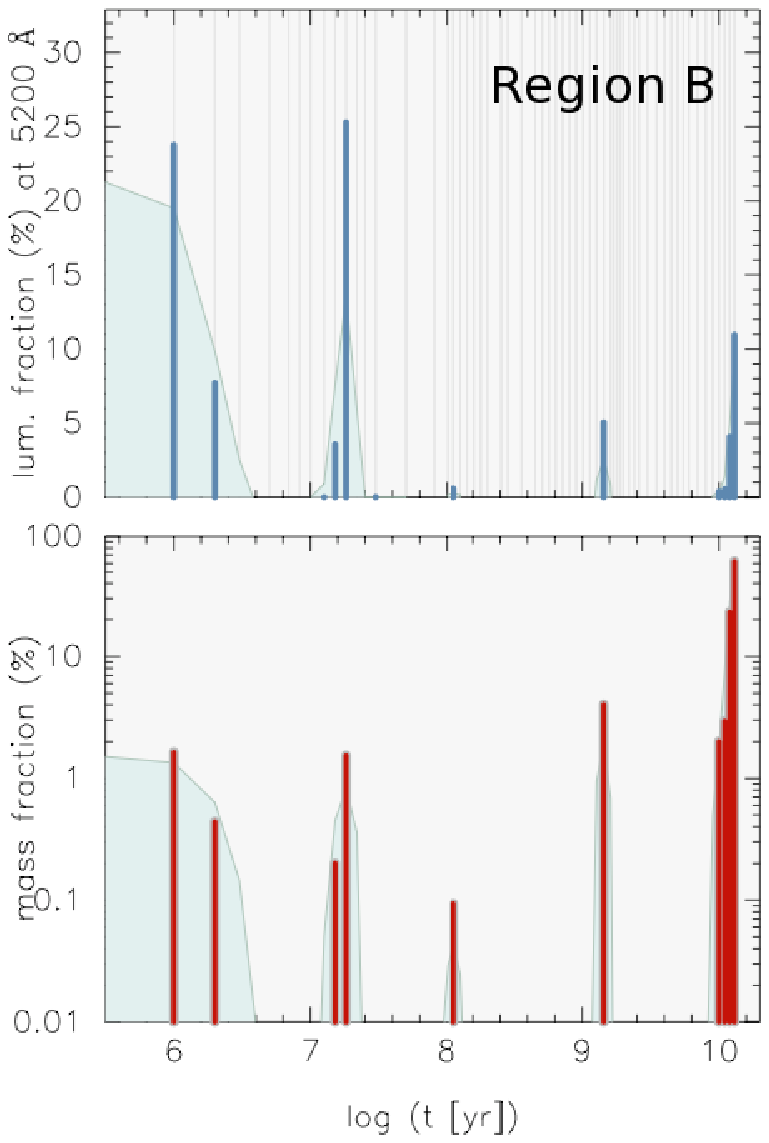}}}
\resizebox{0.33\hsize}{!}{\rotatebox{0}{\includegraphics{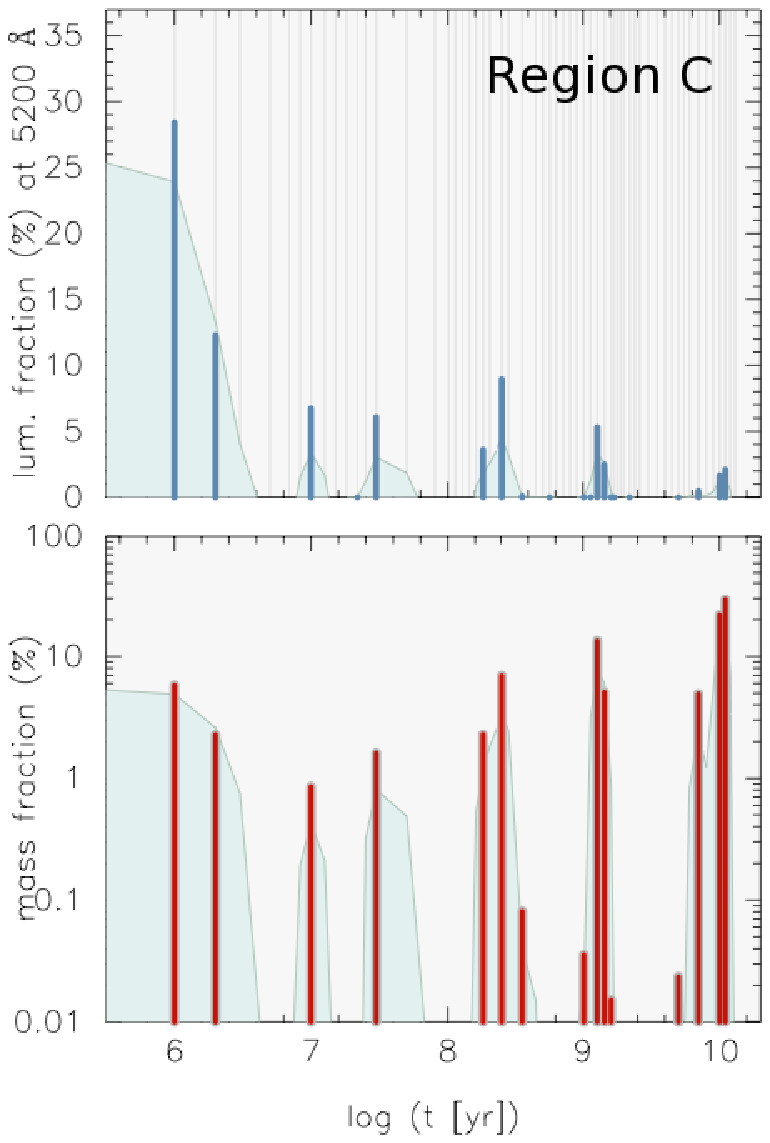}}}
\caption{In the top panels of Figure 15, we show the {\sl Starlight} fit for region B
(the fits to the other two regions are in Figure 1 of the Appendix).
The fit (orange solid line) is fairly good, as shown by the residuals
in the lower panel. A hint to the WR-bump observed by Izotov \& Thuan (1998)
is also visible in the residual emission spectrum above the
best-fitting stellar SED. A comparably good fit has also been
obtained for region C, while for region A, due to its relatively noisy spectrum,
the SFH is more uncertain. The three plots in the bottom of
Figure 15 show the output SFH of the three regions (A in the left, B center, C right), i.e., age distribution of the
SSPs selected by {\sl Starlight} as a function of, respectively, their luminosity
contribution (top insets) to the normalization wavelength and mass contribution (bottom insets).
Thin grey vertical lines in the fraction of luminosity inset indicate the ages of the
library SSPs. The shaded area in the lower diagram shows a smoothed version of
the mass fraction represented by individual SSPs and is meant as schematic illustration of the SFH.}
\label{starfit}
\end{figure*}

From the best-fitting SSP population vector we derived a number of secondary
quantities, of which we present here the luminosity-weighted and mass-weighted stellar age (see Figure~\ref{starfit}).
Prior to spectral fitting, several emission lines and a spectral region around
7500 \AA \, affected by sky line subtraction residuals were interactively
flagged. 

We remark that the input spectra are not corrected for the contribution of
nebular continuum emission, thus introducing some uncertainty on
the outcome of the fitting procedure. Ideally, in galaxies with very strong nebular
emission, the preferable procedure would be to compute and subtract
from the observed spectrum a synthetic nebular line and
continuum spectrum, in order to recover and separately model
the spectral energy distribution (SED) of the stellar background
(see \citealp{1998A&A...338...43P} for an application to the metal-poor
BCG SBS 0335-052E).
However, in the case of Mrk\,930, since the EW(\hb) is
relatively low ($\sim 90$ \AA, \citealp{2000ApJ...531..776G}), and due to the
moderate S/N of our spectroscopic data, especially for region A, we prefer not
to adopt that approach and only flag emission lines.

As spectral synthesis involves several free parameters and the SFH quoted from it is subject to uncertainties, we restrict
ourselves to a qualitative comparison between the main SFH features, as
derived from {\sl Starlight} models, with the reconstructed CFH based on our star
cluster analysis. 
 
In the top panels of Figure~\ref{starfit}, we show an example of the fit produced by {\sl Starlight} (the fits to the others two spectra are in Figure~\ref{starfit-2} of the Appendix). The fit to the observed spectrum of region B (orange solid line) is fairly good, as showed by the residuals in the panel below. An hint of the WR-bump observed by \citet{1998ApJ...500..188I} is also observed in the residual emission spectrum above the best-fitting 
stellar SED. A similar good fit has also been obtained for region C, while the outputs residuals of the Region A  are quite noisy suggesting a slightly poorer fit (see Figure~\ref{starfit-2} of the Appendix). The three plots in the bottom of Figure~\ref{starfit} show the output SFH of the regions as function of the fraction of luminosity (top insets) and stellar mass (bottom insets).

In region B and C, between 76\% and 80\% of the observed flux at 
the normalization wavelength (5200 $\AA$) is due to a young 
stellar population formed within the past $t_{\rm y}=10^8$ yr. 
The mass fraction of these young stars $M_{\rm y,\star}$ is inferred to be
5.5\% in region B and up to $\sim$16\% in region C.

The dominant luminosity contribution of young stars is as well 
reflected upon the low luminosity-weighted stellar age of 
$t_L$=2.1$\pm$1.3 Gyr and 0.6$\pm$0.3 Gyr for regions B and C, respectively. 
This is by a factor of at least 5 lower than the respective mass-weighted 
stellar ages $t_M$ of 11.5$\pm$0.9 Gyr and 6.0$\pm$1.9 Gyr, suggesting, consistently, that old stars dominate 
by mass, more specifically, that 50\% of the total stellar mass was 
already at place 10 Gyr ago.
With regard to the recent SFH of B\&C, the bottom plots in Figure~\ref{starfit} reveal 
at least two salient features: a very recent ($\la$5 Myr) and intense burst of 
star formation (SF) which led to an estimated 4\% and 13\% of the stellar mass 
in regions B and C and a previous episode of strongly enhanced SF between 10 and
$\sim$40 Myr ago. Spectral fits also suggest that region B has been relatively 
quiescent in the age interval between $\sim$50 Myr and 1 Gyr, with merely a
hint for a minor SF event $\sim$100 Myr ago. 
This is not the case, however, for region C which, according to our spectral 
models has undergone significant SF activity between $\sim$200 to 350 Myr ago, 
with the stellar mass formed during that period being comparable to $M_{\rm y,\star}$.

The lower S/N ratio of our spectra for region A does not permit
a conclusive investigation of possible WR features in it.
Additionally, due to the poor seeing during the observations of
that region, it has not been possible to disentangle and separately fit the 
spectrum of its NW and SE part. 
Spectral synthesis to the integral spectrum of region A indicates 
that this region experienced a strong burst of star formation between 25 and
30 Myr ago which produced $\approx$11\% of its stellar mass, followed by a
weaker ($\approx$1\%) SF episode about 5 Myr ago.

\subsection{Comparing star formation and cluster formation history}

Spectral fits consistently imply that a strong burst of star
formation was initiated $\sim$40 Myr ago almost coevally in regions B and C. 
This is in good agreement with the analysis of star clusters 
which reveals a sizeable population in the age interval between 
10 and 40 Myr in either components.
Similarly, spectral synthesis models corroborate the evidence for a 
substantial population of very young ($\la$5 Myr) stars that was already 
drawn from the  analysis.

\citet{2000ApJ...531..776G} estimated from equivalent width EW(\hb) measurements a burst age of 4-5 Myr for Mrk\,930. From the analysis of archival {\it Spitzer} Infrared Spectrograph data,  \citet{2006ApJ...641..795O} found for Mrk\,930 a [Ne\,{\sc iii}]/[Ne\,{\sc ii}] ratio of 3.55. The ratio between these two infrared lines is considered an indicator of the hardness of the UV radiation field (dominated by the short-lived massive stars), and can be converted into an estimate of the burst age. Using the estimates produced by \citet{2006A&A...446..877M} of the [Ne\,{\sc iii}]/[Ne\,{\sc ii}] variation as function of the burst age (Figure 13 in their paper), we find a consistent upper limit for the burst age of 4 Myr.  One should be aware, however, that the line ratios depend on the ionisation parameter, i.e., the ratio of the mean photon flux to the mean atom density. Thus both the radiation field and the pressure of the ambient ISM should be considered. Dynamical modelling of the evolution of H{\sc ii} regions around young clusters show that there is a loose coupling between the age and mass of the cluster and the ionisation parameter which then can be constrained to a rather narrow range \citep{2006ApJ...647..244D}. Using the results from Dopita et el. we find that the age determinations based on the [Ne{\sc iii}]/[Ne{\sc ii}] line ratios should be accurate to within a factor of 2. The age recovered by this line ratio corresponds, however, to a peak in the cluster production, as shown in the previous sections. The agreement among different tracers seams to suggest that the burst has reached a peak roughly 4 Myr ago, or more in general, that it is actively producing stars at the present time.

 There is a considerable amount of gas which can fuel the present burst. However, we are not able to predict whether this burst episode will evolve into a more quiescent phase or will continue converting the gas into stars with a constant rate for the next 1 Gyr.

The {\sl Starlight} outputs are consistent with the stellar
component in Mrk\,930 being predominantly old, with half of the stellar mass
having being assembled $\sim$10 Gyr ago.  
Despite exhibiting a dramatic burst of star formation at the present epoch, Mrk\,930 seems to be an ancient system.

We have not found any very old (10-14 Gyr) cluster in the starburst regions. On the other hand, we do detect some globular clusters with such age range in the galaxy, as showed in the mass-age diagram in Figure~\ref{age-mass}. Their masses, however, are a few per mille of the estimated total stellar mass of the galaxy. These old clusters are in the outskirts of the starburst regions (Figure~\ref{pos-gal}) where an old stellar population has also been revealed (Micheva et al. in prep). Two mechanisms may explain the presence of evolved globular cluster in the outskirts and not in the starburst regions. Firstly, there are old clusters in these regions but they are not detected due to crowding and/or the bright and diffuse nebular emissions. Secondly, the merger event has probably perturbed the dynamical equilibrium of the involved systems. The old galactic stellar population has homogeneously settled   on the new potential, while the old clusters have moved on more distant orbits,  similarly to the old globular clusters observed at high galactic latitude in the Milky Way.

\section{Summary}

We have described the starburst properties of the blue compact galaxy  Mrk\,930 by undertaking a detailed multiwavelength study of its hundreds of resolved star clusters, using imaging from the WFPC2 on board the Hubble Space Telescope, and supported by ground-based narrow-band imaging.

Optical and IR line ratios ([O\,{\sc iii}]/\hb, [Ne\,{\sc iii}]/[Ne\,{\sc ii}]) suggest that the burst age in Mrk\,930 is $\sim 4$ Myr. The age distribution of the star clusters supports this scenario, with a prominent peak in the cluster age distribution between 3 and 4 Myr. The derived cluster formation history shows a rapid rise in the last 10-20 Myr, with a rate at the present time of 1.3 $\msun/yr$ of stars forming in bound star clusters. In total, we find that roughly 25 \% of the star formation is happening in clusters, supporting a scenario where the host environment favours the formation of more numerous and massive clusters. 

The recent burst episodes traced by the analysis of the underlying stellar population are in fairly good agreement with the star formation history reconstructed from the clusters. Young stellar populations dominate the total luminosity of the galaxy. However, the galaxy assembled more than 50 \% of its total stellar mass more than 10 Gyr ago. We detect a few, very old globular clusters in the outskirts of the starburst region, additional support for the  ancient origin of Mrk\,930.

The recovered extinction in the very young clusters shows considerable spread and decreases at older ages. Our analysis suggest that we 
are mainly seeing the optically bright objects, i.e., systems that are still only partially embedded in their natal cocoons. The deeply embedded clusters are most likely too faint to be detected at visible wavelengths. We map the extinction across the galaxy using low-resolution spectra and ground based imaging (\ha/\hb \  ratio)and we compare these results with the extinction distribution derived from the clusters.  We find that the mean optical extinction derived in the starburst regions is very similar to the values typically observed in the clusters (more than 80 \% of systems have $E(B-V) \leq 0.2$ mag), but do not trace clusters locally more extinguished. 

Similarly to other blue compact galaxies like ESO\,185-IG13 and Haro\,11, a considerable fraction of clusters in Mrk\,930 shows a flux excess 
at $\lambda \geq 0.8$ $\mu$m, evident  from our $I$ and $H$ band photometry.  The origin of the excess remains unknown, but we consider a range of possibilities. We notice a correlation between the $I$ band excess and the age of the clusters ($< 6$ Myr). The extended red emission remains the most plausible mechanism which may cause the $I$-band excess. The longer wavelength IR excess that shows up in the $H$-band requires a different mechanism.  At young ages it could be produced by hot dust 
and/or a high fraction of Young Stellar Objects (hot dust in circumstellar disks). Both of these components have been observed in  nearby resolved young star clusters. 
At greater ages, a contribution from Red Super Giant stars or other more "exotic" mechanisms is possible.

\section*{Acknowledgments}
AA thanks A. Bik for a careful reading of the manuscript. PP would like to thank Roberto Romano and Ricardo Amorin
for their cooperation during the TNG observations. The referee is thanked for the accurate revision of the manuscript. AA, G\"O and EZ acknowledge support from the Swedish Research council  (VR) and the Swedish National Space Board (SNSB). PP is supported by a Ciencia 2008 contract, funded by FCT/MCTES (Portugal) and POPH/FSE (EC). G\"O is a Royal Swedish Academy of Sciences research fellow, supported by a grant from the Knut and Alice Wallenberg foundation. NB acknowledges support from the  Swedish Research Council. RMR acknowledges support from STScI through grant GO-10902. This research has made use of the NASA/IPAC Extragalactic Database (NED) which is operated by the Jet Propulsion Laboratory, California Institute of Technology, under contract with the National Aeronautics and Space Administration.

\appendix

\begin{figure*}
\resizebox{0.8\hsize}{!}{\rotatebox{0}{\includegraphics{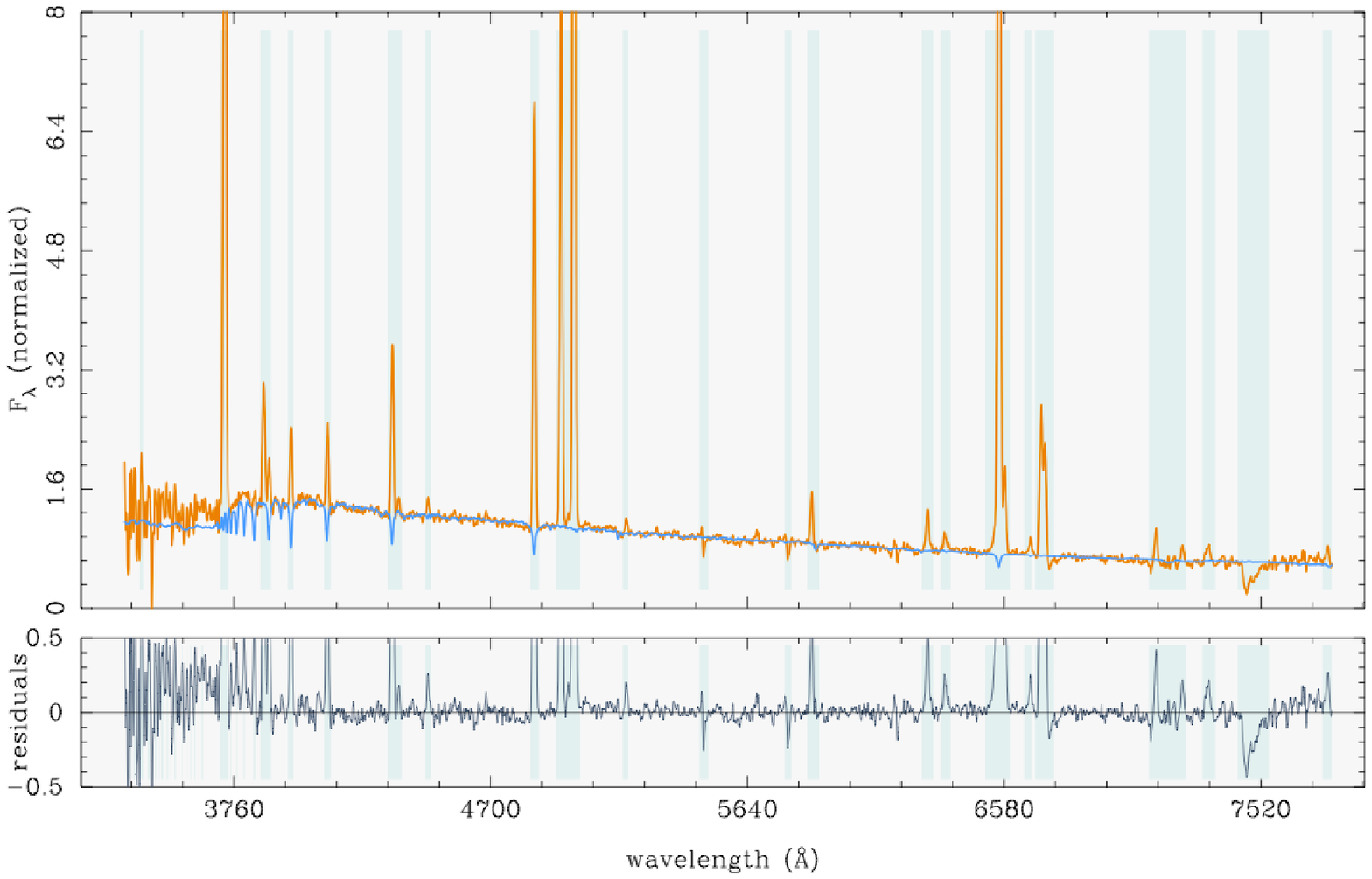}}}
\resizebox{0.8\hsize}{!}{\rotatebox{0}{\includegraphics{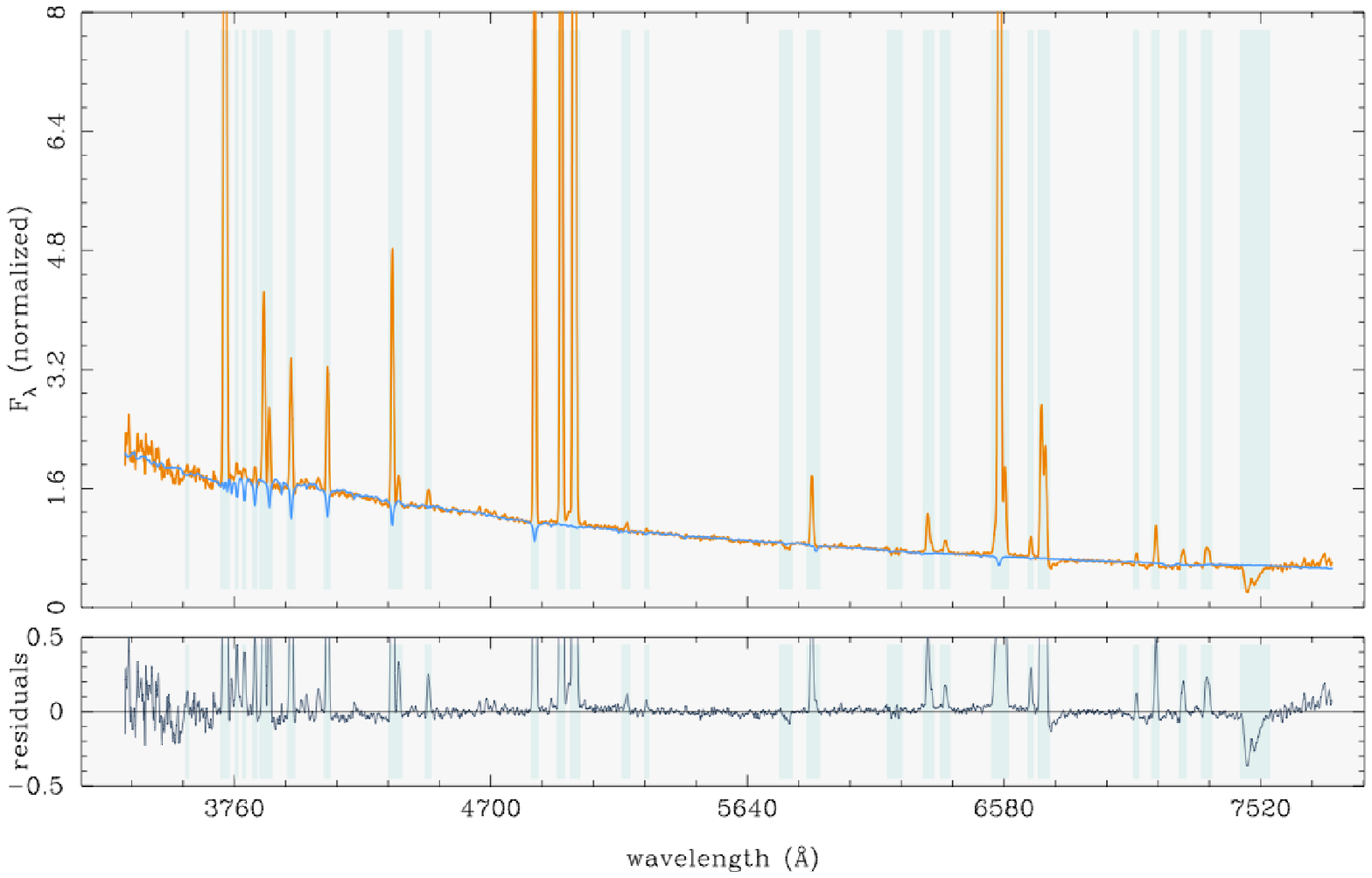}}}
\caption{Continue from Figure~\ref{starfit}. {\sl Starlight} fit to the observed spectra (orange solid line), normalized 
at 5200 $\AA$, of knot A (upper panel) and region C (lover one). Superimposed is the best-fitting stellar SED (blue solid line). The corresponding residuals are displayed below each panel. The shadowed regions of the spectrum have been excluded from the fit.}
\label{starfit-2}
\end{figure*}

\bsp

\label{lastpage}

\end{document}